\documentclass[sigconf, nonacm]{acmart}

\usepackage{amsmath,amsfonts}

\usepackage{amssymb}
\usepackage{graphicx}
\usepackage{textcomp}
\usepackage{xcolor}
\usepackage{xspace}
\usepackage{graphics}
\usepackage{epsfig}
\usepackage{enumitem}
\newcommand{\Paragraph} [1] {\smallskip\noindent{\bf #1 }}
\usepackage{booktabs}
\usepackage{multirow}
\usepackage{bbding}
\usepackage{url}
\usepackage{tikz}
\usepackage{filecontents}
\usepackage{pifont}
\usepackage{graphicx}
\usepackage{balance}  % for  \balance command ON LAST PAGE  (only there!)
\usepackage{subfig}
\usepackage{verbatim}
\usepackage{color,xcolor}
\usepackage{array}
\usepackage{bbding}
\usepackage{amsmath}
\usepackage{algorithm}
\usepackage[noend]{algpseudocode}
\usepackage{algpseudocode}
\newcommand\vldbdoi{10.14778/3705829.3705837}
\newcommand\vldbpages{173 - 186}
\newcommand\vldbvolume{18}
\newcommand\vldbissue{2}
\newcommand\vldbyear{2024}
\newcommand\vldbtitle{\shorttitle} 
\newcommand\vldbavailabilityurl{https://github.com/AiX-im/NeutronTP}
\newcommand\vldbpagestyle{empty}

\newcommand{\notecolor}[1]{\textcolor{black}{#1}}
\newcommand{\system}{NeutronTP\xspace}

\begin{document}
\title{NeutronTP: Load-Balanced Distributed Full-Graph GNN Training with Tensor Parallelism}

\settopmatter{authorsperrow=5}
\author{Xin Ai}
\affiliation{%
  \institution{Northeastern University\country{China} \\ aixin0@\\stumail.neu.edu.cn}
}
% \email{chenchaoy@stumail.neu.edu.cn}
\author{Hao Yuan}
\affiliation{%
  \institution{Northeastern University\country{China} \\ yuanhao@\\stumail.neu.edu.cn}
}
% % \email{gaodechao@stumail.neu.edu.cn}
\author{Zeyu Ling}
\affiliation{%
  \institution{Northeastern University\country{China} \\ lingzeyu@\\stumail.neu.edu.cn}
}
\author{Qiange Wang}
\affiliation{
    \institution{National University of Singapore\country{Singapore} \\
    wang.qg@nus.edu.sg}
    }
\author{Yanfeng Zhang}
\affiliation{
  \institution{Northeastern University\country{China} \\ zhangyf@\\mail.neu.edu.cn}
}
% % \email{zhangyf@mail.neu.edu.cn}
% % % \author{Qiange Wang}
% % % \affiliation{%
% % %   \institution{Northeastern University\country{China\\wangqiange@\\stumail.neu.edu.cn}}
% % % }
% % %\email{wangqiange@stumail.neu.edu.cn}
\author{Zhenbo Fu}
\affiliation{%
  \institution{Northeastern University\country{China} \\ fuzhenbo@\\stumail.neu.edu.cn}
}
% \email{fuzhenbo@stumail.neu.edu.cn}
\author{Chaoyi Chen}
\affiliation{%
  \institution{Northeastern University\country{China} \\ chenchaoy@\\stumail.neu.edu.cn}
  % \streetaddress{P.O. Box 1212}
  % \state{Ireland}
}
\author{Yu Gu}
\affiliation{%
  \institution{Northeastern University\country{China} \\ guyu@\\mail.neu.edu.cn}
}
% \email{guyu@mail.neu.edu.cn}
\author{Ge Yu}
\affiliation{%
  \institution{Northeastern University\country{China} \\ yuge@\\mail.neu.edu.cn}
}
% \author{Xin Ai$^{1}$, Hao Yuan$^{1}$, Zeyu Ling$^{1}$, Qiange Wang$^2$, Yanfeng Zhang$^{1}$, Zhenbo Fu$^{1}$, Chaoyi Chen$^{1}$, Yu Gu$^{1}$, Ge Yu$^{1}$}
% \affiliation{%
%   \institution{$^{1}$Northeastern University, China; $^2$National University of Singapore, Singapore. }}
% \email{{aixin0,yuanhao,lingzeyu,fuzhenbo,chenchaoy}@stumail.neu.edu.cn, {wang.qg@nus.edu.sg}}
% \email{{zhangyf, guyu,yuge}@mail.neu.edu.cn}

% 中
%%
%% The "author" command and its associated commands are used to define the authors and their affiliations.
% \affiliation{%
%   \state{Ireland}

% \orcid{0000-0002-1825-0097}
% \affiliation{%
%   \country{Iceland}


\begin{abstract}

% Full-graph training on graph neural networks (GNNs) has emerged as a promising training method for its effectiveness. Full-graph training requires extensive memory and computation resources. Thus, researchers have proposed employing distributed training to address these resource-intensive requirements. 
% \zyf{polish}
%Distributed GNN training has emerged as a promising direction to address the resource-intensive requirements of large-scale GNN training.
Graph neural networks (GNNs) have emerged as a promising direction. Training large-scale graphs that relies on distributed computing power poses new challenges.
Existing distributed GNN systems leverage data parallelism by partitioning the input graph and distributing it to multiple workers.
However, due to the irregular nature of the graph structure, existing distributed approaches suffer from unbalanced workloads and high overhead in managing cross-worker vertex dependencies. 
%主要针对负载均衡和扩展性，节省通信量不那么容易说明。

%然而，由于图数据不规则的特性,现有的分布式方法承受着负载不均与管理跨worker顶点依赖的高额开销。

% However, due to extensive cross-worker vertex dependencies in distributed GNN training, existing distributed approaches suffer from unbalanced workloads and poor scalability. 
%然而，由于图数据天然的复杂性，现有GNN数据并行因为划分图数据而承受不平衡的工作负载和差的扩展性
%然而GNN张量并行也面临着诸多挑战，因为在GNN模型的每一层计算中，通常包含一些不能使用feature切片进行部分计算的非线性变化（NN操作）。这不仅造成频繁的全局通信去收集与切分特征，还影响着GNN张量并行对于复杂GNN模型的通用性。因此，我们提出了一种解耦的GNN张量并行训练方法，将NN操作与图聚合操作分离，避免二者紧耦合执行造成的频繁通信以及通用性问题。
%GNN张量并行拥有完全平衡的计算与通信负载，无需额外的分区和调度开销。然而，GNN张量并行面临着显著的通信瓶颈，因为每次NN运算都需要使用完整的特征，这就需要频繁的全局通信来连接与切分特征。因此，我们将NN运算与图聚合运算解耦，将全局通信限制在连续多次NN运算的开始和结束位置，同时减少通信量与通信频率。另外，我们还涉及了

%我们提出了TenparaGNN，一个可以确保完全负载均衡的分布式系统。TenparaGNN 利用 GNN 张量并行性，每个工作者负责特定特征维度的完整图GNN训练，从而消除了分布式训练中的跨工作者顶点依赖。GNN张量并行通过划分规则的特征而不是复杂的图结构实现了良好平衡计算与通信。然而，GNN张量并行面临着频繁的全体通信因为NN操作中非线性函数无法部分计算。因此，我们提出了一种解耦的GNN张量并行训练方法，将NN操作与图聚合操作分离，将全体通信限制在连续多次NN运算的开始和结束位置。此外，TenparaGNN提出了节省内存的子图调度策略去支持超过单个GPU内存的大规模图训练，同时重叠子图的通信与计算任务进一步提升性能。

%在本文中，我们扩展张量并行到分布式GNN训练中，通过切分规则的特征而不是复杂的图结构来消除跨worker的顶点依赖。GNN张量并行中的不同worker将负责相同维度的特征切片的GNN训练，实现完全的负载均衡。此外，我们还通过两个关键功能实现高效的GNN张量并行。首先，我们提出了通用的解耦训练框架来将NN操作与图聚合操作解耦，显著降低因为NN操作无法使用特征切片部分计算而造成的通信开销。其次，我们提出了内存高效的子图调度策略去支持超过单个GPU内存的大规模图训练，同时重叠子图的通信与计算任务进一步提升性能。

%\zyf{which is for tensor parallelism, but too far}
In this paper, we leverage tensor parallelism for distributed GNN training. GNN tensor parallelism eliminates cross-worker vertex dependencies by partitioning features instead of graph structures. Different workers are assigned training tasks on different feature slices with the same dimensional size, leading to a complete load balance. We achieve efficient GNN tensor parallelism through two critical functions. Firstly, we employ a generalized decoupled training framework to decouple NN operations from graph aggregation operations, significantly reducing the communication overhead caused by NN operations which must be computed using complete features. Secondly, we employ a memory-efficient task scheduling strategy to support the training of large graphs exceeding single GPU memory, while further improving performance by overlapping communication and computation. By integrating the above techniques, we propose a distributed GNN training system \system. Our experimental results on a 16-node Aliyun cluster demonstrate that \system achieves 1.29$\times$-8.72$\times$ speedup over state-of-the-art GNN systems including DistDGL, NeutronStar, and Sancus.

\end{abstract}

\maketitle
\thispagestyle{empty}
% \vspace{-0.08in}
\pagestyle{\vldbpagestyle}
\begingroup\small\noindent\raggedright\textbf{PVLDB Reference Format:}\\
Xin Ai, Hao Yuan, Zeyu Ling, Qiange Wang, Yanfeng Zhang, Zhenbo Fu, Chaoyi Chen, Yu Gu, Ge Yu. \vldbtitle. PVLDB, \vldbvolume(\vldbissue): \vldbpages, \vldbyear.\\
\href{https://doi.org/\vldbdoi}{doi:\vldbdoi}
\endgroup
\begingroup
\renewcommand\thefootnote{}\footnote{\noindent
This work is licensed under the Creative Commons BY-NC-ND 4.0 International License. Visit \url{https://creativecommons.org/licenses/by-nc-nd/4.0/} to view a copy of this license. For any use beyond those covered by this license, obtain permission by emailing \href{mailto:info@vldb.org}{info@vldb.org}. Copyright is held by the owner/author(s). Publication rights licensed to the VLDB Endowment. \\
\raggedright Proceedings of the VLDB Endowment, Vol. \vldbvolume, No. \vldbissue\ %
ISSN 2150-8097. \\
\href{https://doi.org/\vldbdoi}{doi:\vldbdoi} \\
}\addtocounter{footnote}{-1}\endgroup

% \vspace{-0.05in}
\ifdefempty{\vldbavailabilityurl}{}{
\vspace{.3cm}
\begingroup\small\noindent\raggedright\textbf{PVLDB Artifact Availability:}\\
The source code, data, and/or other artifacts have been made available at \url{\vldbavailabilityurl}.
\endgroup
}

%实验计划
%our system:TensplitGNN   
%comparison system:Neutronstar, G3(未开源), SANCUS
%现有分布式训练系统受限于顶点依赖，产生一系列问题
% (1) 计算与通信负载不均衡
% (2) depcacahe 冗余计算   depcomm产生大量通信，分布式通信一直是分布式计算的主要瓶颈
% 宏观看分布式计算，主要解决的问题 一个是负载均衡，一个是通信瓶颈，进一步看分布式GNN训练，同样受制于这两个问题，并且问题的主要原因就是大量顶点依赖。
%1.motivation实验
%使用系统:NeutronStar, TensplitGNN
%数据集：Reddit   算法：GCN 
%环境：2-layar GCN： 2-node， 4-node， 8-node; 
%     3-layar GCN： 2-node， 4-node， 8-node；
%     4-layar GCN： 2-node， 4-node， 8-node；
%扩展集群机器数与加深模型深度会加剧顶点依赖，我们要通过变换模型深度和机器数来观察这一点
% (1) 测量Nts depcache下的总冗余计算量和每个worker的计算量，观察每个worker是否负载不均
% (2) 测量Nts depcomm下的总通信量和每个worker的通信量与计算量，观察每个worker是否负载不均
% (3) 测量TensplitGNN的总通信量和计算量（可以直接算出来），证明负载均衡
% 上述测量可以模拟
% (4) 对比具体执行时间，说明负载均衡的优势，记录总时间，通信时间，计算时间。

%2. 性能对比试验
%使用系统:NeutronStar, SANCUS, TensplitGNN
%数据集：Reddit, Ogbn-products, Amazon, Ogbn-papers   算法：GCN
%环境：2-layar GCN： 2-node， 4-node， 8-node;
%     3-layar GCN： 2-node， 4-node， 8-node;
% 对比具体执行时间，说明负载均衡的优势，记录总时间，通信时间，计算时间。

\section{Introduction}

\begin{figure}
  \centering
  \includegraphics[width=2.3in]{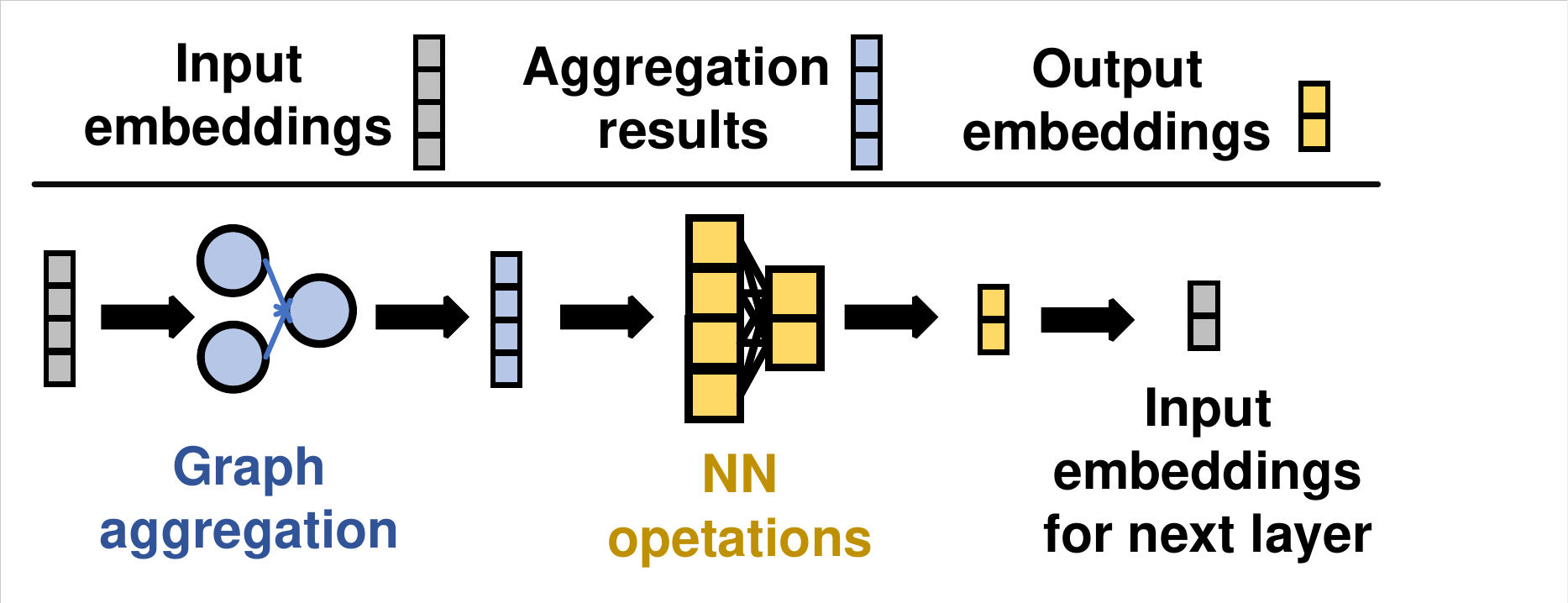}
  \vspace{-0.1in}
  \caption{Illustration of a single-layer computation process in a GNN model, including graph aggregation operations and neural network (NN) operations.}
  \label{fig:GNN_example}
  \vspace{-0.2in}
\end{figure}

%Recently, full-graph GNN training that trains on the entire graph, has emerged as a promising GNN training method for its effectiveness brought by full-neighbor aggregation semantic and full-batch gradient descent \cite{}. However, full-graph training requires high computing power. The training involves random vertex data access and neural network computation, requiring high memory bandwidth and massively parallel computation.

%图结构数据是许多真实世界应用的自然表达，例如社交网络和知识图谱。最近的工作扩展了传统DNN去捕获图数据中的结构信息。这种被称为图神经网络的新型DNN在节点分类和推荐系统等机器学习任务中实现了最佳的性能。最近，在整张图上训练的Full-graph GNN训练因其全邻居聚合语义和全批次梯度下降带来的有效性而成为一种很有前途的GNN训练方法。考虑到从应用中生成的图规模巨大，我们必须进行大规模的并行和分布式计算来处理GNN。

%GNN具有高表达能力的原因在于，GNN从数据样本之间的关系中学习，而传统的DNN是在没有结构信息的单个样本上进行训练的。图A显示了GNN的计算过程，其中包括图聚合操作和NN操作：在每个GNN层中，获得一个顶点的新嵌入首先需要聚合其上一层的邻居嵌入，然后应用NN操作。通过迭代的执行这两步操作，GNN模型可以获得多跳邻居的结构信息。

% Graph-structured data naturally represents various real-world applications such as social networks and knowledge graphs. 
% Recent efforts \cite{Graphsage_2017,GCN_iclr_2017,GAT_ICLR_2018,GNNpower_iclr_2019} have extended traditional deep neural networks (DNNs) to mine structural information within graph data. 
% This novel type of DNNs, known as
Graph Neural Networks (GNNs) have demonstrated remarkable effectiveness in machine learning tasks \cite{simplegcn_icml_2019,textclass_aaai_2019, WEB_KDD_2018,rec_www_2019}. 
Graph-structured data serves as the input for GNNs, where each vertex is associated with a high-dimensional feature vector.
The expressive power of GNNs stems from their ability to learn from relationships between data samples, whereas traditional DNNs are trained on individual samples \cite{GNNlayer-2020}. Figure \ref{fig:GNN_example} illustrates the computational process of GNNs, involving graph aggregation and neural network (NN) operations. In each GNN layer, obtaining a vertex's new embedding entails aggregating its neighbors' embeddings from the previous layer (or neighbors' features at layer 0) and then applying NN operations. By iteratively performing these two steps, the GNN model can capture structural information from multi-hop neighbors.

\begin{figure}
  \centering
  \includegraphics[width=\linewidth]{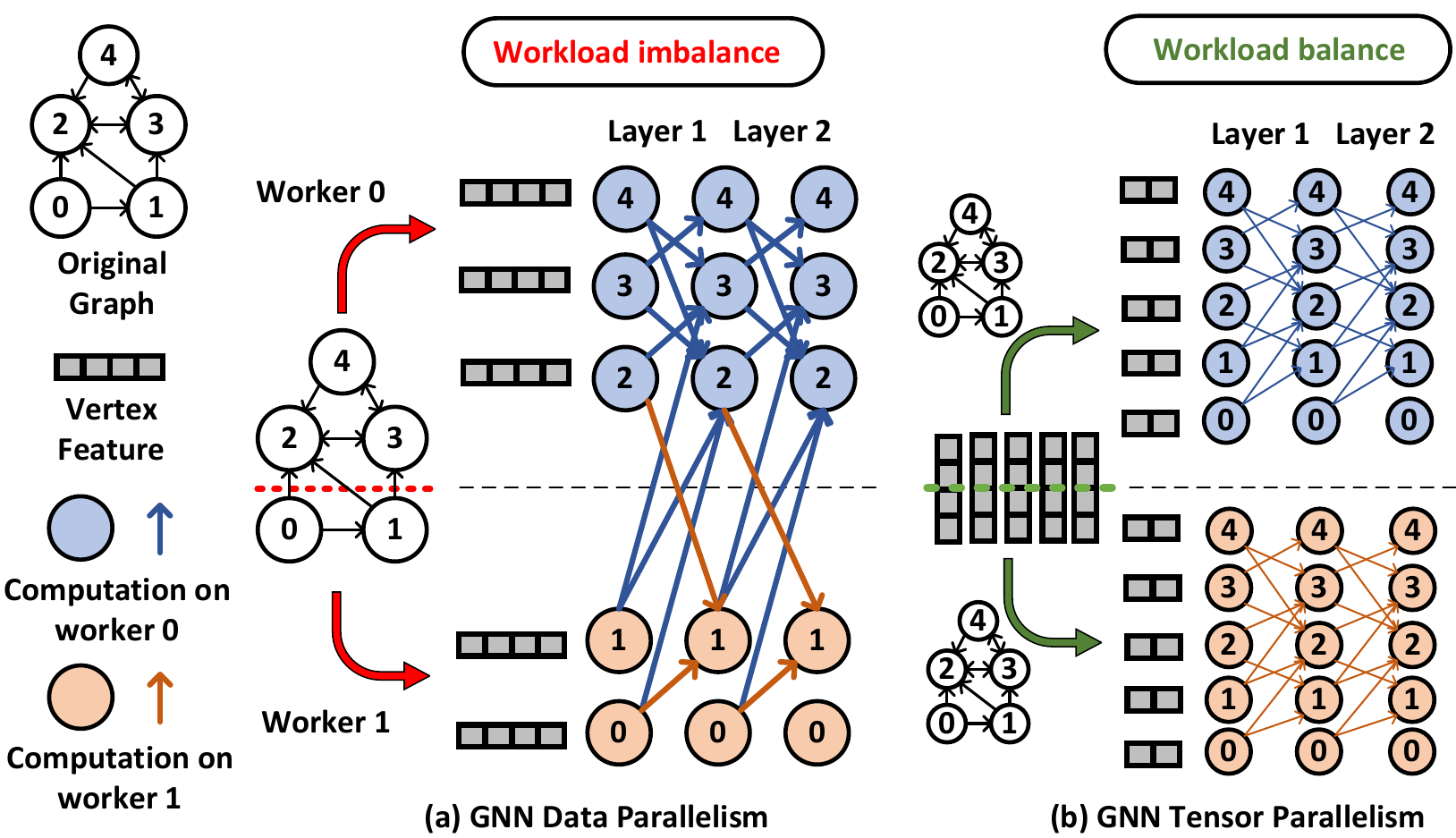}
  \vspace{-0.1in}
  \caption{GNN data parallelism vs. GNN tensor parallelism. The thickness of the arrows and the size of the circles are positively proportional to the feature/embedding dimension and indicate the computation volume of GNN training.}
  \label{fig:motivation}
  \vspace{-0.1in}
\end{figure}

Recently, full-graph GNN training, which involves training on the entire graph, has emerged as a promising GNN training method for its effectiveness brought by full-neighbor aggregation semantics and full-batch gradient descent \cite{ROC_mlsys_2020,neutron_sigmod_2022,distgnn_sc_2021,hongtu_SIGMOD_2024,G3_SIGMOD_2023}. Given the massive scale of graphs generated from applications, large-scale parallel and distributed computing becomes imperative for handling GNNs effectively \cite{distgnn_survey_1,distgnn_survey_2,exp_vldb_2024}. A common approach to scaling GNN training on large-scale graph data is data parallelism, where the graph data is partitioned across different workers for parallel training \cite{aligraph_vldb_2019,distdgl_sc_2020,neutron_sigmod_2022,ROC_mlsys_2020,SANCUS_VLDB_2022,distgnn_sc_2021,AGL_VLDB_2020,BNSGCN_MLSYS_2022,G3_SIGMOD_2023,hongtu_SIGMOD_2024,CAGNET_2020_SC}. 
Despite that partitioning graph data enables distributed GNN systems to handle large-scale data, it also constitutes a primary constraint on the performance of GNN data parallelism. Firstly, as illustrated in Figure \ref{fig:motivation} (a), the irregular nature of graph data makes it challenging to ensure load balance when partitioning the workload. Many survey papers \cite{distgnn_survey_1,distgnn_survey_2,exp_vldb_2024} highlight workload imbalance as a primary challenge, both in mini-batch and full-graph training. Secondly, the edges among data samples (i.e., vertices) lead to complex cross-worker vertex dependencies since graph aggregation may require neighbor data located on remote workers \cite{ROC_mlsys_2020,neutron_sigmod_2022}.
Existing systems adopt methods such as cross-worker neighbor replication \cite{aligraph_vldb_2019,distdgl_sc_2020,AGL_VLDB_2020,P3_OSDI_2021} and neighbor communication \cite{ROC_mlsys_2020,SANCUS_VLDB_2022,G3_SIGMOD_2023,BNSGCN_MLSYS_2022,distgnn_sc_2021} to manage vertex dependencies. As a result, the efficiency of GNN data parallelism is constrained by redundant computations and substantial communication overhead \cite{neutron_sigmod_2022,G3_SIGMOD_2023,ROC_mlsys_2020}.

In this paper, we leverage tensor parallelism for distributed GNN training, eliminating cross-worker vertex dependencies by partitioning features instead of the graph structure. GNN tensor parallelism efficiently balances workload by evenly partitioning vertex features along dimensions. As illustrated in Figure \ref{fig:motivation} (b), GNN tensor parallelism divides vertex features according to the number of workers. Different workers are responsible for GNN training with feature slices of the same dimension, achieving complete computational load balance. GNN tensor parallelism involves two communication operations: \texttt{gather} and \texttt{split}. They collect complete embeddings for NN operations at each model layer, as NN operations include some non-linear operations that cannot be partially computed, and then redistribute the embedding slices back to the corresponding workers. These two communication operations involve all vertices. We only need to ensure that each worker handles the communication task of the same number of vertices to achieve load-balanced communication.

We further enhance the efficiency of GNN tensor parallelism by optimizing communication and memory overhead. Firstly, we employ a generalized decoupled training approach to reduce communication overhead, avoiding frequent execution of two communication operations in each model layer. 
Inspired by existing decoupled GNN training methods \cite{PPRGO_KDD_2020,DAGNN_KDD_2020,DDYG_vldb_2023}, we decouple NN operations from graph aggregation operations, confining \texttt{split} and \texttt{gather} operations to occur before and after consecutive graph operations, significantly reducing communication overhead. 
Additionally, we support decoupled training of complex models \cite{GAT_ICLR_2018} through the precomputation of edge attention, providing generalized support for decoupled training. 
Secondly, we employ a memory-efficient task scheduling strategy to reduce memory overhead, mitigating out-of-memory errors caused by loading the entire graph topology during training. This strategy offers a lightweight subgraph logical partitioning method and further enhances performance by overlapping the computation and communication of different subgraphs.

By integrating the above techniques, we propose \system, a distributed GNN training system that achieves a well-balanced workload. We make the following contributions in this paper. 

%1.我们提出了一种基于tensor parallelism的分布式GNN训练方法，消除了GNN data parallelism中的跨worker顶点依赖，实现了完全的负载均衡。
%2.我们提出了一种通用的解耦训练方法来将NN操作与图聚合操作解耦，显著减少了GNN tensor parallelism中的通信量与通信频率，在集群规模与模型深度上升时获得了良好的扩展性。
%3.我们开发了用于全图GNN训练的分布式系统Tensplit，它拥有完全均衡的工作负载且不需要额外的划分与调度开销，同时重叠了通信与计算过程，从而实现了高性能。

%1.我们对现有 GNN data parallelism方法的性能和局限性进行了全面研究，并发现了大量交错交叉工作顶点依赖性这一关键问题。
%2.我们利用张量并行性来消除分布式 GNN 训练中的大量顶点依赖性，并采用解耦训练方法来显著降低全局通信频率。
%3.我们同时从 GNN 模型中解耦顶点相关和边缘相关的 NNN 操作，为各种输入 GNN 模型提供解耦训练。

\begin{itemize}[leftmargin=*]
    % \item \textbf{Providing insights into the existing GNN data parallelism approaches.} We conduct a comprehensive study on the performance and limits of the existing GNN data parallelism approaches and identify the key issue about the large and interleaved cross-worker vertex dependencies. 
    
    % \item \textbf{Proposing a GNN tensor parallelism method.} We employ tensor parallelism to mitigate the substantial vertex dependencies in distributed GNN training and adopt a decoupled training approach to significantly reduce the frequency of global communication.
    
    % \item \textbf{Proposing a generalized decoupled training method.} We concurrently decouple vertex-associated and edge-associated NN operations from the GNN model, offering decoupled training for various input GNN models.
    \item We propose a distributed GNN training method based on tensor parallelism, which eliminates cross-worker vertex dependencies and achieves complete load balancing. 

    \item We propose a generalized decoupling training method to separate NN operations from graph aggregation, significantly reducing communication frequency in GNN tensor parallelism.
    %, and achieving good scalability as cluster size and model depth increase

    \item We propose a memory-efficient task scheduling strategy to support large-scale graph processing and overlap the communication and computation.
    
    \item We develop \system, a distributed system for full-graph GNN training that utilizes tensor parallelism to achieve fully balanced workloads and integrates a series of optimizations to achieve high performance.

\end{itemize}

% We evaluate \system on a 16-node Aliyun GPU cluster. The experimental results show that \system outperforms the state-of-the-art GNN systems, i.e., 1.29$\times$-6.36$\times$ speedups over DistDGL \cite{distdgl_sc_2020}, 6.15$\times$ speedups over DistDGLv2 \cite{dglv2_kdd_2022}, 4.68$\times$-8.72$\times$ speedups over NeutronStar \cite{neutron_sigmod_2022}, and 3.41$\times$-4.81$\times$ speedups over SANCUS \cite{SANCUS_VLDB_2022}.

We evaluate \system on a 16-node Aliyun GPU cluster. The experimental results show that \system outperforms the state-of-the-art GNN systems on homogeneous graphs, achieving 1.29$\times$-6.36$\times$ speedups over DistDGL \cite{distdgl_sc_2020}, 4.68$\times$-8.72$\times$ speedups over NeutronStar \cite{neutron_sigmod_2022}, and 3.41$\times$-4.81$\times$ speedups over SANCUS \cite{SANCUS_VLDB_2022}. Additionally, \system achieves 6.15$\times$ speedups over DistDGLv2 \cite{dglv2_kdd_2022} on heterogeneous graphs.

The rest of this work is organized as follows. Section \ref{sec2} describes the background and motivations. Section \ref{sec3} provides a detailed description of the proposed GNN tensor parallelism. Section 4 gives an overview of \system and describes the generalized decoupling training method and the memory-efficient task scheduling strategy. 
Section \ref{sec5} presents results. Section \ref{sec6} presents a discussion on related work. Section \ref{sec7} concludes the paper.

% Some systems such as DGL \cite{} and P3 \cite{} use sampling-based training methods to mitigate the overhead by sampling only a small part of the graph during each training iteration. However, sampling operations can generate biased results \cite{} and the duplicate work still exists and grows exponentially to the number of GNN layers \cite{}.

\section{Background and Motivation}
\label{sec2}

\subsection{Graph Neural Networks}

%图结构数据是图神经网络（GNN）的输入，其中每个顶点或边都与高维特征向量相关联。典型的GNN模型由多层组成，为每个顶点计算低维嵌入。这些嵌入可用于执行各种任务，如节点分类和链接预测。每一层GNN模型都包含一个聚合阶段和一个更新阶段。例如，在一个有 $L$ 层的 GNN 中，在 $l$ 层的聚合阶段，每个顶点 $v$ 将其在 $l - 1$ 层的邻居嵌入向量与自己的嵌入向量相结合，使用聚合函数生成聚合结果 $a_{v}^l$

Graph-structured data is input to GNNs, with each vertex having a high-dimensional feature vector.
A typical GNN computes low-dimensional embeddings for vertices through multiple layers, aiding tasks like node classification and link prediction.
Each layer includes an aggregation and an update phase \cite{GNNlayer-2020}.  
In a GNN with $L$ layers, during layer $l$'s aggregation phase, each vertex $v$ aggregates its neighbors' embeddings from layer $l-1$ and its own to produce $a_{v}^l$ using an $AGG$ function:
\begin{align}
    a_{v}^{l} &=AGG(h_u^{l-1} | \forall {u} \in {N_{in}(v)} \cup  \{v\})
 \label{form:agg_forward}
\end{align} 
\noindent where ${N_{in}(v)}$ represents the incoming neighbors of vertex $v$, $h_v^{l}$ represents the embedding vector of vertex $v$ at $l$-th layer, and $h_v^0$ is the input feature of vertex $v$. Next, during the update phase, each vertex computes its output embedding vector $h_{v}^l$ by applying an $UPDATE$ function to the aggregation result $a_{v}^l$:
\begin{align}
    h_v^{l}&=UPDATE(W_v^{l}, a_{v}^{l})
 \label{form:update_forward}
\end{align}
After $L$ layers, each vertex’s feature vector becomes a low-dimensional embedding of its neighbors up to $L$ hops away. 

Both the $AGG$ and $UPDATE$ functions can be neural networks, which are updated during training.
For simple GNN models, such as GCN \cite{GCN_iclr_2017}, which only incorporate vertex-associated NN operations. The computational formulas for GCN are as follows:

\begin{align}
     AGG: a_{v}^{l} =\sum_{u\in N_{in}(v) } (\frac{1}{\sqrt{deg_{in}(v) \cdot deg_{out}(u)} } \cdot h_{u}^{l-1} )
 \label{form:GCN_agg}
\end{align} 

\begin{align}
     UPDATE: h_{v}^{l} = \sigma ( W^{l}a_{v}^{l})
 \label{form:GCN_update}
\end{align}

\noindent, where $deg_{out}(u)$ is the out-degree of vertex $u$, and $deg_{in}(v)$ denotes the in-degree of vertex of vertex $v$. 
The update phase applies standard DNN operations, including a matrix multiplication and a ReLU activation function (i.e., $\sigma$) to the aggregation result.

For complex GNN models, such as GAT \cite{GAT_ICLR_2018}, which include edge-associated NN operations and vertex-associated NN operations. The computational formulas for GAT are as follows:

\begin{align}
AGG:\begin{cases}
    a_{uv}^{l} = softmax(\widehat{\sigma}(a^{T} [W^{l}h_{u}^{l-1} ||W^{l}h_{v}^{l-1}] ))
    \\
    a_{v}^{l} =\sum_{u\in N_{in}(v) } (a_{uv}^{l} h_{u}^{l-1})
\end{cases}
\end{align} 

% \begin{gather}
%     AGG: a_{uv}^{l} = softmax(\widehat{\sigma}(a^{T} [W^{l}h_{u}^{l-1} ||W^{l}h_{v}^{l-1}] ))
%     \\
%     a_{v}^{l} =\sum_{u\in N_{in}(v) } (a_{uv}^{l} h_{u}^{l-1})
%  \label{form:GAT_agg}
% \end{gather} 

\begin{align}
     UPDATE: h_{v}^{l} = \sigma ( W^{l}a_{v}^{l})
 \label{form:GAT_update}
\end{align} 

\noindent The aggregate phase first needs to assign an edge weight $a_{uv}$ for each incoming edge to vertex $v$. This process involves concatenating (i.e., $[\cdot||\cdot]$) and mapping (i.e., $a^{T}$) the parameterized representations of the source $u$ and destination $v$ to derive edge-wise attention coefficients. Then, these coefficients are fed to a LeakyReLU activation function (i.e., $\widehat{\sigma}$) and use a softmax function to compute normalized edge weight for subsequent neighborhood aggregation. The update phase is the same as GCN. 

%对于简单GNN模型，例如GCN，GraphSAGE，他们只在UPDATE阶段包含顶点相关的NN操作。他们的计算公式如下：

%对于复杂GNN模型，例如GAT，将在aggregate阶段包含边相关的NN操作和在update阶段包含顶点相关NN操作。他的计算公式如下
%聚合阶段首先需要为每个顶点的每条入边指派一个边权a_{ij}。这一过程通过对源顶点与目的顶点的特征进行拼接与映射得到边相关注意力系数。然后再通过一个LeakyReLU激活函数与softmax函数得到归一化的边权。

\begin{figure}
  \centering
  \vspace{-0.1in}
  \includegraphics[width=\linewidth]{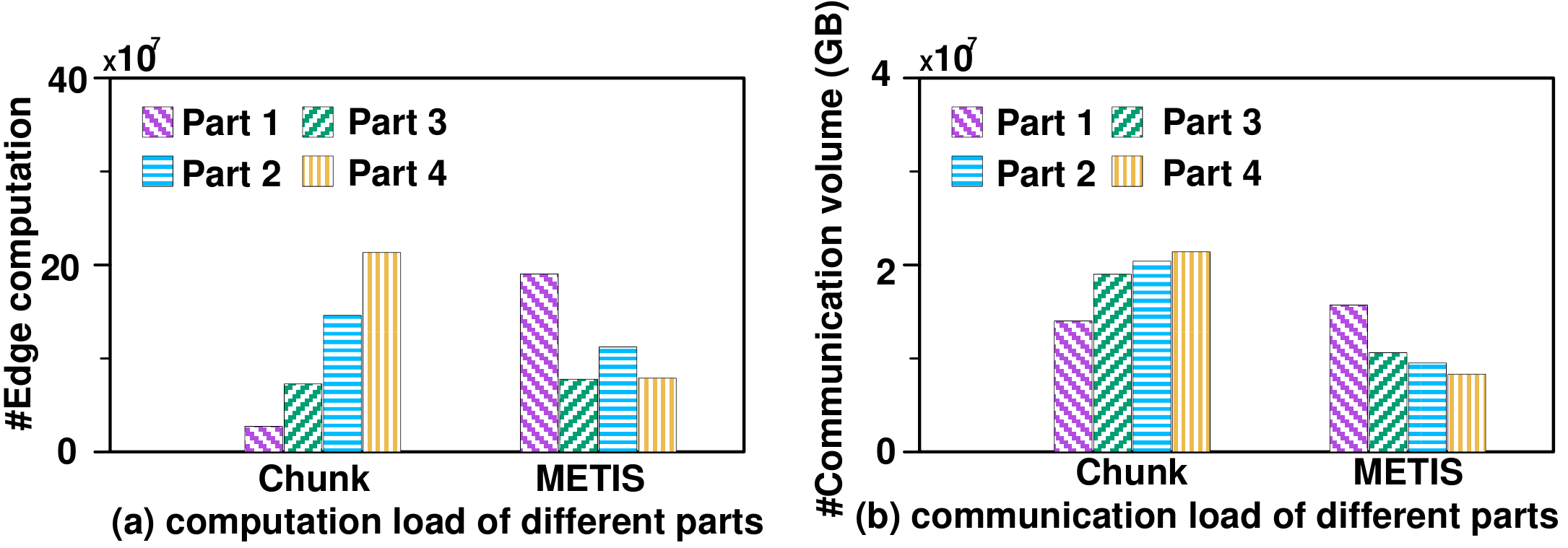}
  \vspace{-0.1in}
  \caption{GNN training workload of 4 partitions under different partitioning methods. (2-layer GCN on Reddit)}
  \label{fig:workload}
    \vspace{-0.2in}
\end{figure}

\subsection{Distributed GNN Training with Data Parallelism}
When dealing with large-scale graphs, single machines' limited memory and computational resources become bottlenecks for large-scale GNN training. Distributed computing offers sufficient computational resources, thereby enhancing training efficiency. 
Existing GNN systems \cite{neutron_sigmod_2022,hongtu_SIGMOD_2024, G3_SIGMOD_2023, ROC_mlsys_2020,neugraph_atc_2019,distdgl_sc_2020,aligraph_vldb_2019} leverage data parallelism by partitioning the input graph and distributing it to multiple workers to train the same GNN model collaboratively. However, due to the graph aggregation operations in GNNs, which create vertex dependencies across these partitions, these graph partitions cannot be processed independently.
In NeutronStar \cite{neutron_sigmod_2022}, the authors summarize the current GNN systems into two categories according to the way they manage vertex dependencies: Dependency Cache (DepCache) methods and Dependency Communication (DepComm) methods. The GNN systems \cite{aligraph_vldb_2019,distdgl_sc_2020,AGL_VLDB_2020,P3_OSDI_2021} employ the DepCache methods to replicate data of neighboring nodes from partitions outside to the local worker, enabling independent GNN training locally but incurring redundant computations.
In contrast, systems \cite{ROC_mlsys_2020,G3_SIGMOD_2023,SANCUS_VLDB_2022,distgnn_sc_2021,BNSGCN_MLSYS_2022} employing the DepComm methods collect data of neighboring vertices through communication from remote workers. While avoiding redundant computations, these systems incur necessary communication costs. 
The irregular nature of the graph causes extensive cross-worker vertex dependencies in data parallelism and thereby increases the difficulty of high-performance distributed GNN training. We summarize two major limitations below.

\begin{figure}
\vspace{-0.1in}
  \centering
  \includegraphics[width=\linewidth]{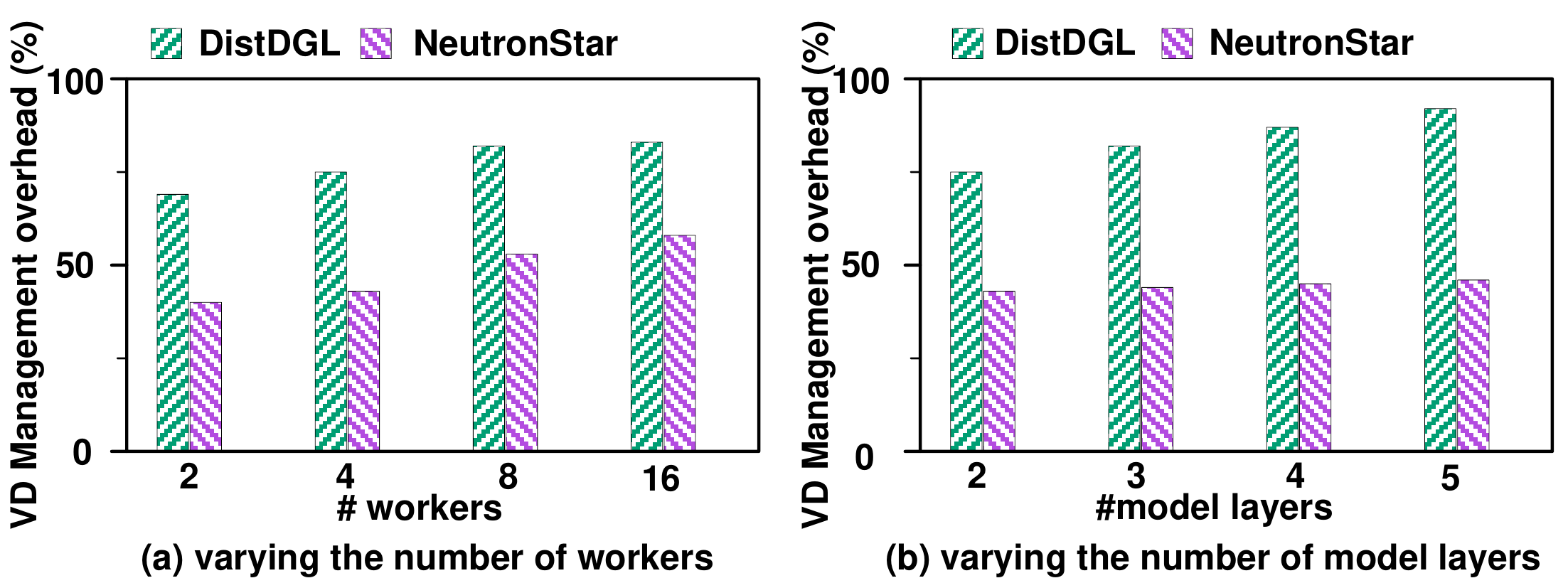}
  \vspace{-0.2in}
  \caption{The VD management overhead of DistDGL and NeutronStar.}
  \label{fig:VD_overhead}
  \vspace{-0.2in}
\end{figure}

\begin{figure}
% \vspace{-0.2in}
  \centering
  \includegraphics[width=\linewidth]{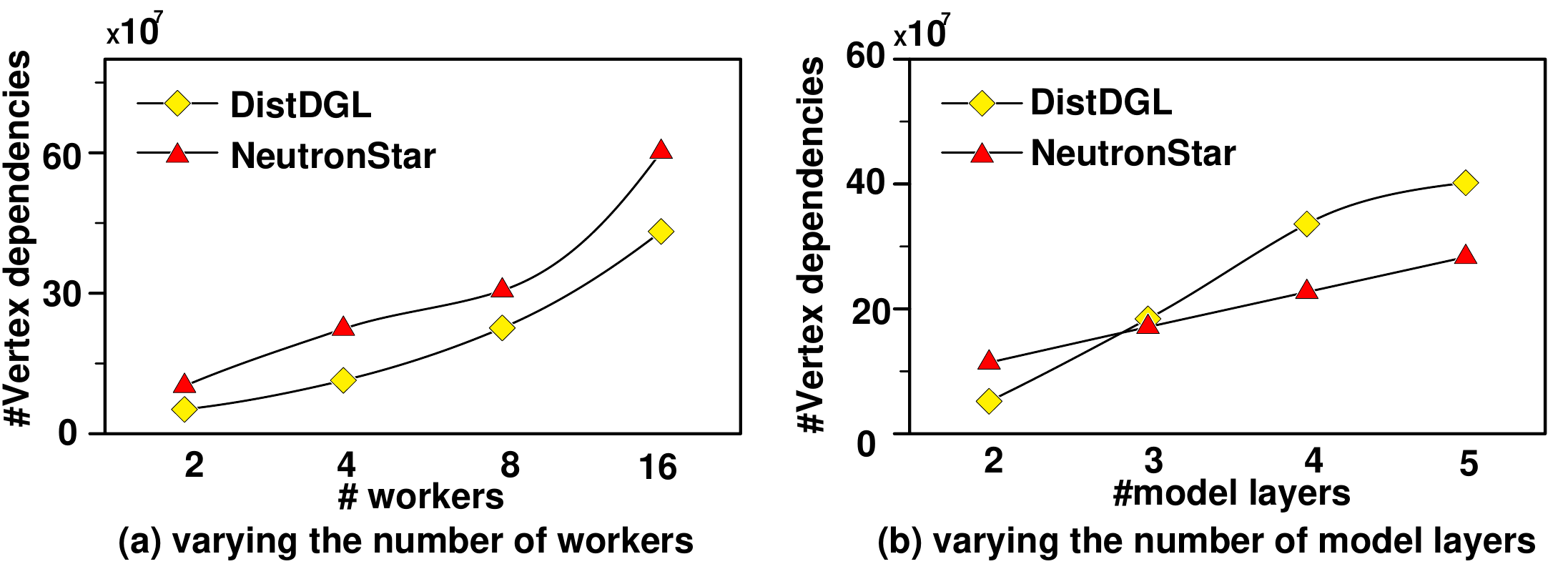}
  \vspace{-0.2in}
  \caption{The number of vertex dependencie of DistDGL and NeutronStar.}
  \label{fig:VD_scale}
  \vspace{-0.2in}
\end{figure}
\Paragraph{Limitation \#1:Workload imbalance.}
Distributed GNN training with data parallelism is prone to workload imbalance due to the skewed interconnection structure of graphs.
To confirm our analysis, we employ Chunk-based graph partitioning and METIS-based graph partitioning in NeutronStar \cite{neutron_sigmod_2022}, evaluating the balance between computational and communication loads across each partition. The results are shown in Figure \ref{fig:workload}.
The chunk-based graph partitioning strategy, employed by systems such as NeuGraph \cite{neugraph_atc_2019}, ROC \cite{ROC_mlsys_2020}, and NeutronStar \cite{neutron_sigmod_2022}, divides the graph into chunks where nodes are arranged with consecutive IDs. Although this method achieves vertex balance, it may lead to significant workload imbalance because it does not account for the edge distribution among workers. On the other hand, the METIS algorithm, utilized by DistDGL \cite{distdgl_sc_2020}, SANCUS \cite{SANCUS_VLDB_2022}, and BNS-GCN \cite{BNSGCN_MLSYS_2022}, aims to minimize cross-worker edges (i.e., edge-cuts) in its partitioning decisions. However, focusing on minimizing edge cuts does not guarantee balanced remote and local vertices within each partition, resulting in varied communication and computation loads across workers.

%管理顶点依赖是现有基于数据并行的分布式训练系统的主要开销。另外，当集群规模与模型深度加深时，管理顶点依赖的开销占比会进一步提升。为了验证这一观点，我们分析现有系统在改变集群规模或模型深度下管理顶点开销的开销占比。我们通过统计通信时间与冗余计算时间来统计管理顶点依赖的开销，并统计通信边与冗余计算边来量化顶点依赖规模。结果如图4所示。在不同集群规模与模型深度下，DistDGL与NeutronStar中管理顶点依赖的开销分别平均占总运行时间的80.6%与46.5%。当集群规模从2扩展到16，DistDGL与NeutronStar中管理顶点依赖的时间占比分别提升了1.21倍与1.45倍，顶点依赖规模分别增加了8.1倍与6.2倍。当模型深度从2加深到5，DistDGL与NeutronStar中管理顶点依赖的时间占比分别提升了1.22倍与1.06倍，顶点依赖规模分别增加了7.7倍与3.0倍。总而言之，GNN数据并行为管理顶点依赖进行了大量通信与冗余计算。同时，无论是因为集群规模扩大而划分更多子图，还是计算图因为模型加深而扩大，都会导致顶点依赖规模显著增加，从而导致更大开销。

%现有基于Data parallelism的GNN训练系统在集群规模
%为了验证这一观点，我们分析现有系统在改变集群规模或模型深度下性能的变化情况，并通过计算通信边与冗余计算边来统计顶点依赖规模。
%如图figure 3 (b)所示，这是因为在模型深度没有改变的情况下，DistDGL与NeutronStar的顶点依赖规模分别增加了8.1倍与6.2倍，降低了集群规模增加带来的好处。
%另一方面，如图figure 3 (c)所示，当模型层数逐渐加深时，顶点依赖也会迅速上升，阻碍现有GNN系统对深层模型的扩展
\Paragraph{Limitation \#2:High overhead in managing cross-worker vertex dependencies.}
Managing vertex dependency (VD) constitutes the primary overhead in GNN data parallelism. Furthermore, as the cluster scale expands or model layers deepen, the proportion of VD management overhead further increases. To confirm our analysis, we analyze VD management overhead in DistDGL \cite{distdgl_sc_2020} and NeutronStar \cite{neutron_sigmod_2022} when altering the cluster scale or model layer. We measure the proportion of VD management overhead by accounting for communication time and redundant computation time and quantify VD's scale by calculating communication and redundant computation edges. 
As shown in Figure \ref{fig:VD_overhead}, across different cluster sizes and model layers, the VD management overhead in DistDGL and NeutronStar averages 80.6\% and 46.5\% of the total execution time, respectively. 
Furthermore, as shown in Figure \ref{fig:VD_scale}, no matter increasing the number of partitions and workers or deepening models, the VD scale can be substantially increased.
When the number of workers scales from 2 to 16, the time proportion of VD management overhead in DistDGL and NeutronStar increases by 1.21$\times$ and 1.45$\times$, respectively, with the VD scale increasing by 8.1$\times$ and 6.2$\times$, respectively. Similarly, as increases the model layer from 2 to 5, the time proportion of VD management overhead in DistDGL and NeutronStar increases by 1.22$\times$ and 1.06$\times$, respectively, with the VD scale increasing by 7.7$\times$ and 3.0$\times$, respectively.
% In summary, GNN data parallelism entails significant communication and redundant computation for managing vertex dependencies. 

% As the cluster scale expands or model layers deepen, there is an observable trend of increasing vertex dependencies. This trend results in additional redundant computations and heightened communication overhead, thereby impeding the scalability of existing GNN systems. 
% To confirm our analysis, we conduct experiments to analyze the efficiency of existing systems when altering the cluster scale or model depth. We quantify the scale of vertex dependencies by calculating communication edges and redundant computation edges. 
% Figure \ref{fig:poor_scala} (a) shows the speedup of per-epoch running time of DistDGL \cite{distdgl_sc_2020} without neighbor sampling and NeutronStar \cite{neutron_sigmod_2022}. When the number of workers scales from 2 to 16, the performance of DistDGL and NeutronStar only increases as much as 1.4$\times$ and 3.8$\times$, far lower than the ideal 8$\times$ speedup ratio, as represented by Linear. This is due to an increase in vertex dependency scale by 8.1$\times$ for DistDGL and 6.2$\times$ for NeutronStar (Figure \ref{fig:poor_scala} (b)), thereby diminishing the benefits of an enlarged cluster scale. 
% On the other hand, as depicted in Figure \ref{fig:poor_scala} (c), with the gradual increase in model depth, the vertex dependencies also increase rapidly, impeding the scalability of existing GNN systems to deeper models.

\subsection{Opportunity: Tensor Parallelism}

%与 GNN 数据并行不同，GNN 张量并行分割的是特征而不是图，每个GPU负责特定特征维度的完整图 GNN 训练。
%避免在不同worker间划分图结构是避免section 2.2中强调的局限的关键。因此，我们提出利用张量并行性进行分布式 GNN 训练，划分顶点特征而不是图结构，消除跨worker的顶点依赖同时保证负载均衡。

%消除数据并行在划分图数据时产生的顶点依赖是避免section 2.2中强调的局限的关键。因此，我们提出利用张量并行性进行分布式 GNN 训练，划分规则的顶点特征而不是复杂的图结构。

% 受到DNN张量并行的启发，该方法通过划分模型参数而不是数据样本来支持大规模模型训练。我们扩展张量并行到分布式GNN训练并将划分对象从模型参数转为顶点特征,因为GNN训练的内存开销主要来自顶点数据（特征和嵌入），模型数据通常很小。传统DNN采用多维划分，例如2D,2.5D,3D划分来减少张量并行的通信开销与内存开销。这些多维划分方法都进一步的将DNN的数据样本切分为多个不相交的子集来与模型参数子集进行矩阵运算。然而，由于GNN独特的图聚合训练语义，进一步切分GNN的数据样本即图数据将会重新引入顶点依赖。这使我们无法直接应用过去的DNN张量并行方法来优化GNN张量并行的效率。在本文中，我们专注于如何设计高效的GNN张量并行，并解释它的优势与挑战。

Avoiding partitioning the graph structure across different workers is crucial to mitigating the limitations outlined above.
Therefore, we exploit tensor parallelism for distributed GNN training, partitioning vertex features instead of the graph structure, thereby eliminating cross-worker vertex dependencies while ensuring load balancing.
Inspired by DNN tensor parallelism \cite{tensorpara_arxiv_2019} that supports DNN training with large-scale model data by partitioning model parameters instead of data samples. We extend tensor parallelism to distributed GNN training and change the partitioning target from model parameters to vertex data, as the memory overhead of GNN training mainly stems from vertex data (i.e., features and embeddings), while model data is typically small. Past DNN works employ multidimensional partitioning, such as 2D \cite{2Dtensor_ipdps_2021}, 2.5D \cite{25Dtensor_2021}, and 3D \cite{3Dtensor_2021} partitioning, to reduce the communication and memory overhead in tensor parallelism. These multidimensional partitioning methods further partition DNN data samples into multiple disjoint subsets for matrix operations with model parameter subsets. 
However, due to the unique graph aggregation training semantics of GNNs, further partitioning GNN data samples (i.e., graph data) reintroduces vertex dependencies. This renders us unable to directly apply past DNN tensor parallelism methods to optimize the efficiency of GNN tensor parallelism.

% In this paper, we focus on designing efficient GNN tensor parallelism and elucidate its advantages and challenges.

%相反，我们探索支持全训练流程的张量并行，在全部模型层中垂直切分feature与嵌入，实现完全的负载均衡。同时，GNN张量并行不再产生

%我们注意到一些最近的研究也在分区之间切分feature。然而，他们仅仅利用这种数据划分方式来减少特征通信开销。我们探索支持全训练流程的张量并行，同时切分特征与嵌入，实现完全的负载均衡并提出新的挑战。P3使用特征切片完成第一次图聚合操作，通过减少通信顶点来降低特征拉取开销。Du et al.在一些迭代中跳过特征拉取操作，仅使用部分特征维度完成本地训练，在模型准确率与系统性能间找到一个平衡。这种feature-based的划分方式不能保证负载均衡因为每个worker内的子图规模并不一致。我们目标在于充分探索full-graph GNN训练中最大程度的tensor并行。

\Paragraph{Recent works in vertical feature partitioning.}
%讨论P3和infocomm24这两篇
%最近一些
%1.针对训练不同 2.只在底层切分feature而不是全部训练 3.有准确率损失，无法确保算法收敛 4.因为采样子图拓扑的规模不一致，通信与计算仍可能不均衡
%我们注意到，最近的一些研究也探讨了类似的 GNN 张量并行思想。然而他们只在部分训练过程采用张量并行，而我们探索支持全训练流程的张量并行，实现完全的负载均衡。P3在不同worker间切分feature并在本地完成第一次图聚合操作，减少了分布式mini-batch训练中特征拉取的开销。Du et al.在一些迭代中跳过特征拉取操作，仅适用部分特征维度完成本地训练，在模型准确率与系统性能间找到一个平衡。他们都仅在部分训练（第一层图聚合或部分轮次）中使用特张量并行, 因此不能应用到Tensplit关注的全流程张量并行训练中中。另外，不完全的张量并行不能保证负载均衡因为每个worker内的子图规模并不一致。我们目标在于充分探索full-graph GNN训练中最大程度的tensor并行。
%我们注意到一些最近的研究也在分区之间切分feature。然而，他们仅仅利用这种数据划分方式来减少特征通信开销。我们探索支持全训练流程的张量并行，同时切分特征与嵌入，实现完全的负载均衡并提出新的挑战。P3使用特征切片完成第一次图聚合操作，通过减少通信顶点来降低特征拉取开销。Du et al.在一些迭代中跳过特征拉取操作，仅使用部分特征维度完成本地训练，在模型准确率与系统性能间找到一个平衡。这种feature-based的划分方式不能保证负载均衡因为每个worker内的子图规模并不一致。我们目标在于充分探索full-graph GNN训练中最大程度的tensor并行。
We note that some recent studies \cite{P3_OSDI_2021,info24} explore vertical feature partitioning in distributed GNN training. However, they employ feature partitioning approach only in partial training processes and focus on reducing the feature communication overhead (See Section \ref{sec6} for details).
This feature partitioning approach cannot guarantee load balancing in the end-to-end training because most training still employs data parallelism. In contrast, we explore tensor parallelism throughout the entire training process, vertically partitioning both features and embeddings across all layers, achieving complete load balancing.

\section{GNN Tensor Parallelism}
\label{sec3}

%在这个section，我们详细介绍了GNN张量并行的工作流程并且通过工作负载分析解释了它的优势与挑战。
In this section, we provide a detailed exposition of the workflow involved in GNN tensor parallelism and elucidate its advantages and challenges through workload analysis.

\subsection{GNN Tensor Parallelism Workflow}

\begin{figure}
% \vspace{-0.2in}
  \centering
  \includegraphics[width=8cm,page={1}]{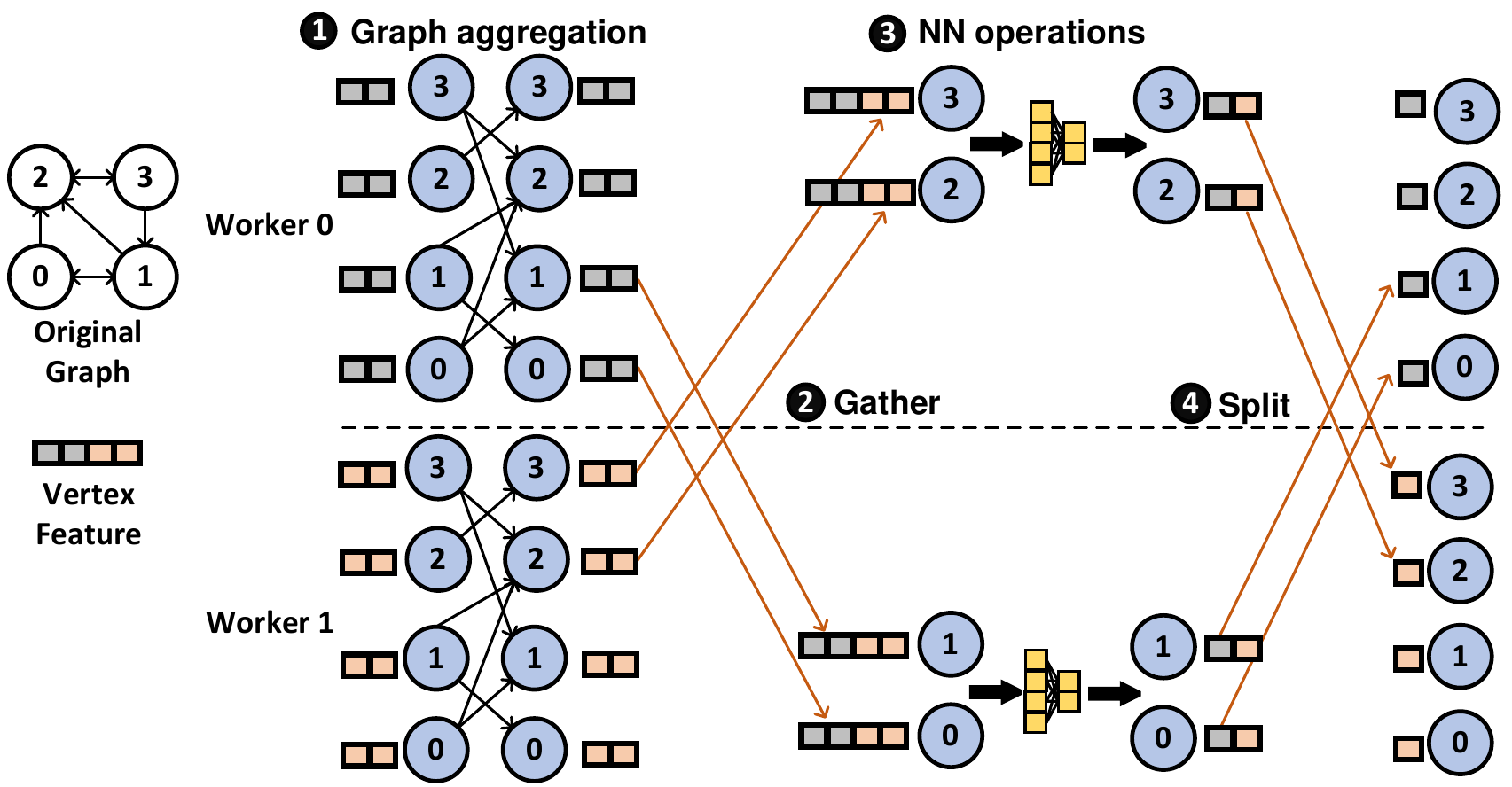}
  \vspace{-0.1in}
  \caption{GNN tensor parallelism workflow for a single layer.}
  \label{fig:workflow}
  \vspace{-0.1in}
\end{figure}

%%%%%%%%%%%%%%%%%%%%%%%%%%%%%% replot
% 现实世界的图数据集通常由表示图结构的邻接矩阵和所有顶点的高维特征矩阵组成。与GNN数据并行划分图拓扑不同，GNN张量并行按照维度对特征矩阵进行划分，每个worker负责完整图的部分特征维度的训练。与GNN数据并行划分图拓扑不同，GNN张量并行按照维度对特征矩阵进行划分，每个worker负责完整图的部分特征维度的训练。图5展示了单层GNN张量并行的训练流程。最初，顶点的特征向量被均匀划分到所有worker中，即每个worker持有所有顶点的特征向量$\frac{D}{N}$维，其中$D$是顶点的总维数 特征向量，$N$是worker的数量。每个worker拥有完整的图结构，并在本地进行全邻居的聚合操作。
% A real-world graph dataset typically consists of the adjacency matrix denoting the graph structure and a high-dimensional feature matrix of all vertices. 
Unlike GNN data parallelism, which partitions the graph topology, GNN tensor parallelism vertically partitions vertex features along dimensions, where each worker is responsible for the complete graph GNN training of different feature slices with the same dimensional size. 
Figure \ref{fig:workflow} illustrates the single-layer training workflow for GNN tensor parallelism. Initially, the feature vectors of the vertices are evenly partitioned among all workers according to their feature dimensions, i.e., each worker holds $\frac{D}{N}$ dimensions of the feature vectors where $D$ is the total number of vertex feature dimensions, and $N$ is the number of workers. \notecolor{The GNN tensor parallelism directly leverages the structure of the raw graph for its computations. Specifically, at each layer of the GNN model, every vertex aggregates information from all its neighboring vertices along the incoming edges and then applies NN operations.}

% 在NN计算开始之前，由于神经网络计算中涉及非线性操作而无法部分计算，因此执行 \texttt{gather} 操作以获得完整的顶点嵌入。此外，为了确保神经网络计算和集体通信任务在不同worker间均匀分配，每个worker负责\frac{V}{N}个顶点的计算和通信，其中V是顶点总数。完成图操作后，所有worker共同进行\texttt{gather}操作，发送各自的嵌入切片给相应的worker。随后，每个worker开始对其负责的顶点进行神经网络计算。在NN计算完成后，所有worker一起执行\texttt{split}操作，按给定维度重新划分局部嵌入，并将其发送回相应的worker，以继续下一层GNN训练。此训练过程在每一层GNN中重复进行。
Each worker stores the complete graph structure and conducts full-neighbor aggregation operations locally (\ding{202}).
Before the start of NN computations, a \texttt{gather} operation is performed to obtain complete vertex embeddings because nonlinear operations in NN computations cannot be partially computed (\ding{203}). To ensure the uniform distribution of NN computation tasks and communication tasks across different workers, each worker is responsible for computing and communicating with $\frac{V}{N}$ vertices, where $V$ represents the total number of vertices and $N$ denotes the number of workers. All workers initiate \texttt{gather} operations simultaneously, sending local embedding slices to the corresponding workers. Subsequently, each worker commences the NN computation tasks for local vertices (\ding{204}). Finally, upon completion of NN computations, all workers initiate \texttt{split} operations simultaneously to re-segment local embeddings and send them to the corresponding workers to continue the next layer of GNN training (\ding{205}).
The \texttt{gather} and \texttt{split} operations are implemented by collective communication libraries such as NCCL \cite{nccl} and Gloo \cite{gloo}.
% This training process repeats in each layer of GNN tensor parallelism.
%%%%%%%%%%%%%%%%%%%%%%%%%%%%%%
% gather 和 split 操作通常由集合通信库实现，如 NCCL 和 Gloo。

\subsection{Workload Analysis}

\newcommand{\timeAgg}{$O(E\frac{D}{N})$\xspace}
\newcommand{\spaceAgg}{$O(V\frac{D}{N})$\xspace}
\newcommand{\timeUpdate}{$O(\frac{V}{N} D^2)$\xspace}
\newcommand{\spaceUpdate}{$O(\frac{V}{N} D)$\xspace}
\newcommand{\spaceGather}{$O(\frac{VD}{N^2})$\xspace}
\newcommand{\timeGatherApprox}{$O(\frac{VD}{N})$\xspace}
\newcommand{\TPCommLoadT}{$N \times 2(N-1)\frac{VD}{N^2}L \approx 2VDL$\xspace}

\newcommand{\TPCommLoad}{$(N-1)\frac{2VD}{N} L \approx 2VDL$\xspace}

\newcommand{\DPCommLoad}{$ {\textstyle \sum_{i=1}^{N}} |R_{i}|DL$\xspace}

\newcommand{\DPAdditionalMemory}{$\sum_{i=1}^{N} (V_i + E_i + \frac{V}{N}D + LD^2) = V+E+VD+NLD^2$\xspace}

\newcommand{\TPAdditionalMemory}{$\sum_{i=1}^{N} (V + E + V\frac{D}{N} + LD^2) = N(V+E) + VD + NLD^2$\xspace}

\notecolor{
% GNN张量并行通过沿维度均匀划分顶点特征，实现了计算和通信负载的均衡。在计算过程中，每个工作节点处理同一维度特征片段的全图聚合操作，并为相同数量的顶点执行神经网络计算。在通信过程中，在\texttt{gather}阶段，每个工作节点从其他$N-1$个工作节点接收本地神经网络计算顶点的嵌入片段。单个工作节点的通信负载可以表示为$\frac{V}{N} \cdot \frac{D}{N} \cdot (N-1)$，其中$V$表示顶点总数，$D$表示特征维数，$N$表示工作节点数。\texttt{split}阶段可以理解为\texttt{gather}阶段的逆过程，每个工作节点的通信负载保持不变，为$\frac{V}{N} \cdot \frac{D}{N} \cdot (N-1)$。
% The time complexity of the graph aggregation operation is \timeAgg, and the space complexity is \spaceAgg. The time complexity of the NN operation is \timeUpdate, and the space complexity is \spaceUpdate, 
GNN tensor parallelism achieves a well-balanced computation and communication load by evenly partitioning vertex features along dimensions. 
In the computation process, each worker handles the full-graph aggregation operations of the same dimension feature slices and performs NN computations for the same number of vertices. The time and space complexities of graph aggregation operation are \timeAgg and \spaceAgg, respectively. The time and space complexities of NN operation are \timeUpdate and \spaceUpdate, respectively, where $V$ and $E$ represent the total number of vertices and edges, $N$ denotes the number of workers, and $D$ denotes the feature dimension, for simplicity, we assume that the feature dimensions are uniform across all layers. In the communication process, during the \texttt{gather} phase, each worker receives the embedding slices of local NN computation vertices from the other $N-1$ workers. The time and space complexities are $(N-1) \cdot \frac{V}{N} \cdot \frac{D}{N} \approx$ \timeGatherApprox and \spaceGather, respectively.
% where \commVolume represents the single communication volume of each worker.
The \texttt{split} phase can be understood as the inverse process of the \texttt{gather} phase, and it has the same time and space complexity as the \texttt{gather} phase.
}

\notecolor{
% 在分布式 GNN 训练中，除了计算和通信之外，每个 worker 还需要额外的内存来存储图数据（即图拓扑和顶点特征）及模型参数。
% GNN 数据并行将图数据划分到多个 worker 上，每个 worker 保存部分图拓扑和对应顶点的特征。因此，GNN 数据并行的额外内存开销为X。
% GNN 张量并行将顶点特征沿特征维度划分到不同的工作器上。每个工作器保存完整图拓扑和所有顶点部分维度的特征。GNN 张量并行的额外内存消耗为 Y。
% In distributed GNN training, each worker needs to store graph data (i.e., graph topology and vertex features) and model parameters. GNN data parallelism partitions the graph data across multiple workers, where each worker stores a portion of the graph topology and corresponding vertex features. In contrast, GNN tensor parallelism partitions vertex features along the dimension, where each worker stores the entire graph topology but only partial dimensions of vertex features. This approach uses more memory to replicate graph structures, eliminating inter-worker vertex dependencies and ensuring load balancing.
% 我们进一步分析了 GNN 张量并行的总计算和通信负载。与单机全图训练相比，GNN 张量并行保持相同的总计算负载，且无任何冗余计算。关于通信负载，GNN 张量并行的总通信负载可表示为 \TPCommLoad。GNN 数据并行的总通信负载可表示为 \DPCommLoad，其中 $R_{i}$ 表示 worker $i$ 的远程顶点。GNN数据并行的总通信量与远程节点的数量有关，随着worker数量的增加，远程节点的数量会显著上升，甚至超过2V个。相比之下，GNN张量并行的总通信量随worker数量增加并无显著变化，因此GNN张量并行的通信负载通常优于数据并行。
We further analyze the total computation and communication load of GNN tensor parallelism. GNN tensor parallelism maintains the same total computational load as single-machine full-graph training without any redundant computations. Regarding communication load, GNN tensor parallelism performs \texttt{split} and \texttt{gather} operations at each layer to communicate embedding slices with the other $(N - 1)$ workers. The total communication load for GNN tensor parallelism is \TPCommLoadT, where $L$ denotes the number of model layers. The total communication load of GNN data parallelism is \DPCommLoad, where $R_{i}$ denotes the remote vertices of worker $i$. As the number of workers increases, the remote vertices ($ {\textstyle \sum_{i=1}^{N}} |R_{i}|$) rise significantly, often exceeding $2V$ \cite{hongtu_SIGMOD_2024}. In contrast, the total communication volume in GNN tensor parallelism remains relatively constant with worker increases, typically having a lower communication load than data parallelism.
}
%%%%%%%%%%%%%%%%%%%%%%%%%%%%%%

\notecolor{In GNN tensor parallelism, more memory is used to replicate the graph structure to eliminate cross-worker vertex dependencies and ensure load balancing. This overhead is generally acceptable since the primary memory consumption in GNN training comes from vertex data, including features, embeddings, and gradients \cite{hongtu_SIGMOD_2024}. For example, in the Ogbn-paper dataset, the graph topology size is 6.4 GB, while vertex features occupy 82.7 GB. GNN tensor parallelism distributes all vertex data across different workers by either dimension or vertex count.}

% The total communication volume of GNN data parallelism is related to the number of remote vertices, however, the number of remote vertices rises significantly as the number of workers increases. In contrast, the total communication volume of GNN tensor parallelism grows linearly with the total number of vertices, making the communication load of tensor parallelism typically better than data parallelism. 
% Nonetheless, as the number of workers increases, the linearly growing communication load may become a potential bottleneck for GNN tensor parallelism. Therefore, reducing the total communication overhead is key to implementing efficient GNN tensor parallelism.
%%%%%%%%%%%%%%%%%%%%%%%%%%%%%%%%%%%%%%%%%%%%%%%%

% \subsection{System Challenges}
% \subsection{Opportunities and Challenges}
\subsection{Challenges}
% GNN 张量并行性的好处也伴随着挑战，必须克服这些挑战才能充分利用加速机会。
% Based on the analysis above, we can observe that GNN tensor parallelism achieves a well-balanced computation and communication load.
The benefits of GNN tensor parallelism come with challenges that must be overcome to fully exploit acceleration opportunities.

% GNN张量并行实现了计算和通信负载的均衡。相较于DNN数据并行，GNN张量并行具有相同的计算负载且没有冗余计算，另外通信负载不会随着分区数量的增多而显著改变。然而，为了构建一个高效的张量并行系统，存在许多重大挑战。
% Based on the analysis above, GNN tensor parallelism achieves a well-balanced computation and communication load. Compared to GNN data parallelism, GNN tensor parallelism maintains the same computational load without any redundant computation. Furthermore, the communication load does not change significantly as the number of workers increases. Nevertheless, building an efficient GNN tensor parallelism system presents many significant challenges.

% Compared to full-graph training on a single machine, GNN tensor parallelism maintains the same total computational load without any redundant computations.

%GNN张量并行带来了巨大的通信瓶颈，因为每个NN操作都需要使用完整的嵌入来保证非线性变换的正确计算。这就需要频繁地进行集体通信，以收集和拆分嵌入式数据。正如我们在3.2中强调的，GNN tensor并行的总通信量可能超过原本的数据并行方式。另外，频繁的全体通信会降低计算效率，因为每一个计算阶段都依赖与上一阶段的通信结果。我们需要进一步探索如何减少整体通信量与通信频率
%To reduce communication overhead, DNN tensor parallelism methods employ multidimensional partitioning, such as 2D \cite{2Dtensor_ipdps_2021}, 2.5D \cite{25Dtensor_2021}, and 3D \cite{3Dtensor_2021} partitioning. However, because GNN tensor parallelism partitions vertex features rather than model parameters, the aforementioned partitioning methods are not applicable to GNN training. 
% As emphasized in Section 3.2, the total communication overhead of GNN tensor parallelism may exceed that of GNN data parallelism. 
%GNN张量并行需要频繁的集合通信，因为每个NN操作都需要使用完整的嵌入来保证非线性变换的正确计算。这需要我们在NN操作前后来收集和拆分嵌入。同时，计算过程与通信过程之间产生大量的layer-wise同步，因为每一个计算阶段都依赖与上一阶段的通信结果。我们需要进一步探索如何减少整体通信量与通信频率
\Paragraph{Challenge \#1: Frequent collective communication.} 
Compared to GNN data parallelism, GNN tensor parallelism involves more rounds of communication (i.e., twice per layer) to gather and split vertex embeddings. This frequent communication may impact computation efficiency due to substantial layer-wise synchronization. Therefore, reducing the overall communication frequency is crucial for effectively implementing GNN tensor parallelism.

\Paragraph{Challenge \#2: Processing the entire graph on a single worker.}
%\wqg{THis name is not intuitive}
% Although the primary memory overhead of GNN training data stems from vertex features and intermediate results rather than the graph structure itself \cite{hongtu_SIGMOD_2024}, 
GNN tensor parallelism becomes impractical when a single GPU memory cannot accommodate the entire graph and corresponding embedding slices.  
To support the training of large-scale graphs, we need to offload the training data to the CPU main memory. This requires the further design of a task scheduling strategy and consideration of integration with pipeline techniques to minimize latency when accessing data in CPU main memory.

\section{The \system}
\label{sec4}

\begin{figure}
% \vspace{-0.2in}
  \centering
  \includegraphics[width=7cm,page={1}]{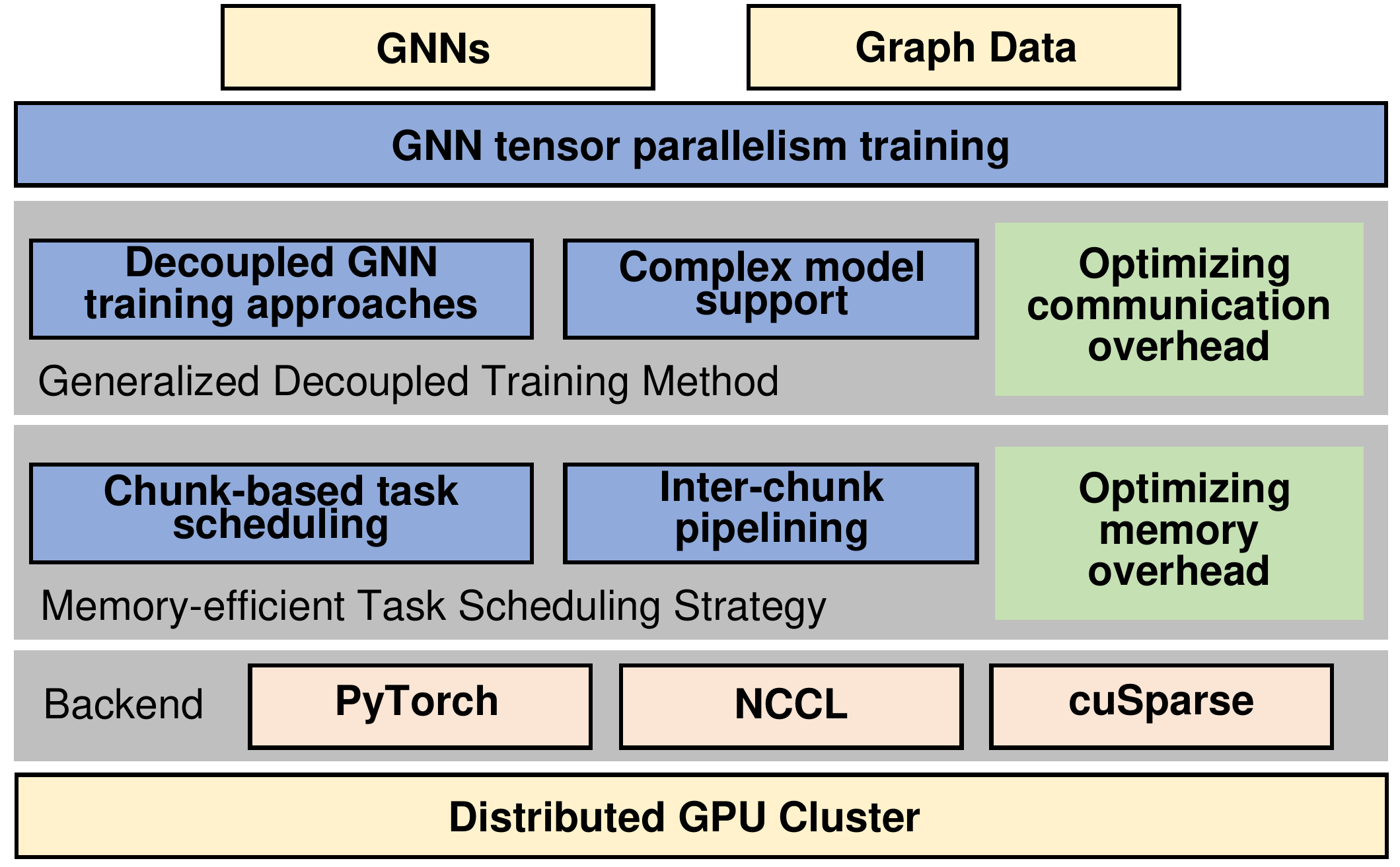}
  \vspace{-0.1in}
  \caption{\system system overview.}
  \label{fig:overview}
  \vspace{-0.1in}
\end{figure}

We present \system, a distributed system for full-graph GNN training that utilizes tensor parallelism and addresses the challenges outlined in Section 3.3 through two critical functions. Figure \ref{fig:overview} provides an architectural overview of \system.

% In each model layer, graph aggregation operations are interleaved with vertex-associated NN operations and may also include edge-associated NN operations. 
% % \sanzo{comment} %这个也是chanllege 2的主要原因吧，

%挑战1的主要原因在于NN操作和图聚合操作的耦合训练模式，这种模式导致GNN张量并行在使用完整嵌入与嵌入切片直接频繁切换。因此，我们将 NN 操作与图聚合操作解耦执行，将集体通信限制在连续图操作的开始和结束，从而减少通信量和通信频率。然而，现有的解耦训练方法并不能支持涉及边相关NN运算的复杂 GNN 模型。为了解决这个问题，我们进一步探索了边缘关联NN运算的解耦方法，提供通用的解耦训练方法。在启动图聚合流程之前，我们使用上一轮的历史嵌入为每条边预先计算与边相关的 NN 运算。
\Paragraph{Generalized decoupled training method. }
The main reason for challenge \#1 lies in the coupled training patterns of NN and graph aggregation operations, which frequently switch between using complete embeddings and embedding slices. 
% \sanzo{, which require frequent exchange of embedding slices.}
Therefore, we decouple NN operations from graph aggregation operations, restricting collective communication to the beginning and end of consecutive graph operations, thereby reducing communication frequency.
% However, existing decoupling training methods \cite{PPRGO_KDD_2020,DAGNN_KDD_2020,DDYG_vldb_2023} do not support complex GNN models involving edge-associated NN operations. To address this, we further explore a decoupling approach for edge-associated NN operations, providing a generalized decoupling training method. 
To address the issue that existing decoupling training methods \cite{PPRGO_KDD_2020,DAGNN_KDD_2020,DDYG_vldb_2023} do not support complex GNN models involving edge-associated NN operations, we further explore a decoupling approach for these operations, providing a generalized decoupling training method. 
Specifically, before initiating the graph aggregation, we precompute edge-associated NN operations for each edge.
% using the historical embeddings from the previous round.

%为了解决挑战2，我们在每个worker内进一步划分子任务，细粒度的执行全图GNN训练。TenparaGNN采用了一个内存高效的调度策略，通过chunk-based子图调度减少运行时内存消耗并通过inter-layer的流水线重叠计算与通信进一步提升性能。具体的，我们在每个worker内将全图逻辑划分为多个可以放入GPU内存的chunk，每个chunk内包含一些具有连续id的顶点以及他们的全部入边。在训练过程中，每个worker按相同顺序调度chunk到GPU，保持GNN tensor并行的负载均衡。在不破坏layer-wise同步屏障的前提下，inter-layer pipeline重叠每一层训练内当前chunk的计算任务与上一个chunk的通信任务。

\Paragraph{Memory-efficient task scheduling strategy. } 
To address challenge \#2, we further partition subtasks within each worker to perform fine-grained GNN training. \system employs a memory-efficient task scheduling strategy that reduces runtime memory consumption through chunk-based task scheduling and further enhances performance by inter-chunk pipeline. Specifically, within each worker, we partition the entire graph logically into multiple chunks that can fit into GPU memory, with each chunk containing a set of vertices with contiguous IDs along with all their incoming edges. During training, each worker schedules chunks onto the GPU in the same order, maintaining load balance for GNN tensor parallelism. Without compromising the layer-wise synchronization barrier, the inter-chunk pipeline overlaps the computation and communication of different chunks.

\subsection{Generalized Decoupled Training Method}
\label{sec4}

\notecolor{GNN tensor parallelism requires frequent collective communication for gathering and splitting embeddings by dimensions between NN computation (requiring the embedding split by vertices) and graph propagation (requiring the embedding split by dimensions), which results in substantial layer-wise data synchronization overhead. To improve communication efficiency for high performance training, NeutronTP employs a decoupled training method that reorganizes the execution order by moving NN computation to the beginning or end of the computation graph. This design reduces the collective communication of switching data organizations.
}

\subsubsection{Decoupled GNN Training Approaches}

Previous works \cite{lightGCN_sigir_2020,simplegcn_icml_2019,deepexp_2022_kdd} have indicated that the expressive power of GNNs originates
from NN operations and graph operations themselves, not their coupling execution. Moreover, the coupling execution of graph aggregation and NN operations may lead to over-smoothing problems when training deep GNN models \cite{deepexp_2022_kdd,appnp_iclr_2019,DAGNN_KDD_2020}. Therefore, some decoupled GNN approaches \cite{simplegcn_icml_2019, sign_arxiv_2020, appnp_iclr_2019,adapro_trans_2021,DAGNN_KDD_2020} advocate separating NN operations from graph aggregation. This decoupling method has been shown to effectively enhance both model accuracy and scalability in deep model training. A recent study \cite{DDYG_vldb_2023} has also applied decoupled training methods to dynamic GNN training, achieving significant scalability and performance.
%%%%%%%%%%%%%%%%%%%%%%%%%%%%%%%%%%%%replot
% 现有解耦训练技术通常只关注于解耦顶点相关NN操作，并不支持解耦边相关NN操作。对于包含边相关NN运算的复杂模型，如图形注意网络（GAT)，图聚合本身会引入无法部分计算的非线性运算。如Section 2.1 中公式 (5)所描述，GAT在训练过程中需要为图中所有边计算注意力系数。
However, existing decoupling training methods \cite{simplegcn_icml_2019, sign_arxiv_2020, appnp_iclr_2019,adapro_trans_2021,DAGNN_KDD_2020} typically focus only on decoupling vertex-associated NN operations and do not support decoupling edge-associated NN operations. For complex models incorporating edge-associated NN operations, such as GAT \cite{GAT_ICLR_2018}, graph aggregation in itself may introduce non-linear operations that cannot be partially computed. 
% As described in Equation (5) in Section 2.1, GAT requires computing attention coefficients for all edges in the graph.
% 我们扩展了解耦GNN训练方法，通过使用历史嵌入来计算本轮训练所需的全部注意力系数。该方法将边相关的神经网络操作从图聚合中解耦，从而支持复杂模型的训练。
%%%%%%%%%%%%%%%%%%%%%%%%%%%%%%%%%%%%%%%%%%

% 我们扩展了解耦 GNN 的方法以支持复杂的模型训练。现有解耦训练技术通常只关注于解耦顶点相关NN操作，并不支持解耦边相关NN操作。对于包含边相关NN运算的复杂模型，如图形注意网络（GAT)，图聚合本身会引入无法部分计算的非线性运算。如Section 2.1 中公式 (5)所描述，GAT在训练过程中需要为图中所有边计算注意力系数，输入数据包括源顶点与目的顶点的嵌入与该层的模型参数。为了避免该计算过程与公式(6)中的图聚合操作的耦合执行，我们提出在图聚合操作开始前，使用上一轮的历史嵌入来计算本轮训练所需的全部注意力系数。
% We extend decoupled GNN training methods to support complex model training. Existing decoupling training methods \cite{simplegcn_icml_2019, sign_arxiv_2020, appnp_iclr_2019,adapro_trans_2021,DAGNN_KDD_2020} typically focus only on decoupling vertex-associated NN operations and do not support decoupling edge-associated NN operations. For complex models incorporating edge-associated NN operations, such as Graph Attention Networks (GAT) \cite{GAT_ICLR_2018}, graph aggregation in itself may introduce non-linear operations that cannot be partially computed. 
% As described in Equation (5) in Section 2.1, GAT requires computing attention coefficients for all edges in the graph, with input data consisting of embeddings of source and destination vertices. To avoid coupling this computation with the graph aggregation operation described in Equation (6), we propose computing all the required attention coefficients for this round of training using historical embeddings from the previous round. 

%%%%%%%%%%%%%%%%%%%%%%%%%%%%%%%%%%%%replot
We extend the decoupled GNN training method by precomputing all the attention coefficients required for each edge. This approach further decouples edge-associated NN operations from graph aggregation, thereby supporting the training of complex models.
% 我们扩展了解耦GNN方法，通过历史嵌入的方法将边上的NN计算从图聚合中解耦出来，以支持具有边上NN计算的复杂模型训练。
% We extend decoupled GNN training by using history embeddings to decouple edge-associated NN operations from graph aggregation to support the training of complex models.
% We extend the decoupled GNN training method by using history embeddings to compute all attention coefficients required for this layer.
%%%%%%%%%%%%%%%%%%%%%%%%%%%%%%%%%%%%%%%%%%
% 具体的，\system将在CPU中保存上一轮迭代中每一层模型的本地顶点嵌入，并在本轮图聚合操作开始前，使用data parallelism的方式为为全部边计算注意力系数。因为边注意系数不可避免的需要完整顶点嵌入，同时其计算过程涉及全部边，我们使用data parallelism的方式，在每个worker上计算其本地顶点的全部入边的注意力系数，在计算完毕后，将注意力系数共享给所有worker，开始tensor parallelism训练。通过上述设计，我们可以在开始图聚合操作前，完成边相关NN计算。
%\system stores the local vertex embeddings from the previous iteration on the CPU,
Specifically, before the graph aggregation operation starts in this round, it computes attention coefficients using data parallelism. Since the computation of edge attention coefficients requires complete vertex embeddings and involves all edges, we employ data parallelism to compute attention coefficients for all incoming edges of local vertices on each worker. After the computation is completed, the attention coefficients are shared among all workers.
For other stages, the approach remains consistent with simple GNN models, allowing the use of GNN tensor parallelism. With the above design, we can perform edge-associated NN computations before initiating the graph aggregation operation.
%%%%%%%%%%%%%%%%%%%%%%%%%%%%%%%%%%%%%

% To address this, we employ a hybrid approach of GNN tensor parallelism and GNN data parallelism to support the training of complex GNN models. Specifically, when computing attention coefficients for each edge in the graph, as complete vertex embeddings are inevitably required, we switch to GNN data parallelism at this stage. For other stages, the approach remains consistent with simple GNN models, allowing the use of GNN tensor parallelism. 

% Several methods \cite{simplegcn_icml_2019, sign_arxiv_2020} execute graph operations in advance, and then feed the propagated features into multiple NN operations. on the contrary, several methods \cite{appnp_iclr_2019,adapro_trans_2021,DAGNN_KDD_2020} first transform the vertex features and then propagate the vertex information to distant neighbors. 

% However, existing decoupling methods \cite{PPRGO_KDD_2020, DAGNN_KDD_2020, DDYG_vldb_2023} do not adequately support complex GNN models involving edge-associated NN operations. To address this, we explore a decoupling approach for edge-associated NN operations, treating multi-hop edges spanning all model layers as a hyper-edge from the outermost source vertices to the innermost destination vertices. We perform a single edge-associated NN operation on this hyper-edge, decoupling it from multiple consecutive graph aggregation operations. We also provide theoretical analyses demonstrating that a single edge-associated NN operation on the hyper-edge can approximate multiple NN operations on multi-hop edges and ensure model convergence.

\begin{figure}
% \vspace{-0.2in}
  \centering
  \includegraphics[width=0.95\linewidth]{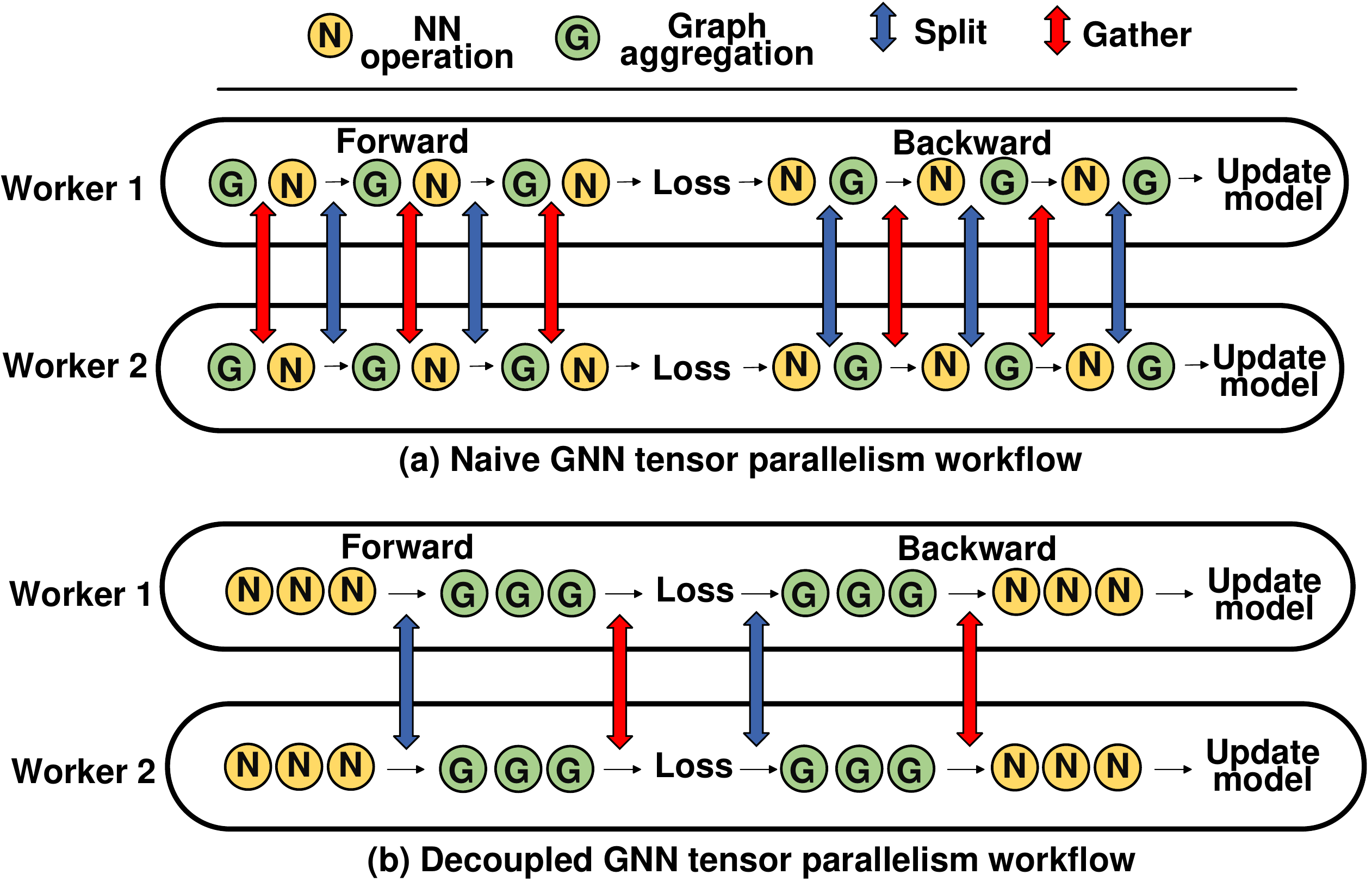}
  \vspace{-0.15in}
  \caption{An illustrative example for showing communication frequency of naive GNN tensor parallelism and decoupled GNN tensor parallelism (3-layer GNN).}
  \label{fig:DGTP}
  \vspace{-0.2in}
\end{figure}

% \zyf{move this sentence to text, For any number of model layers, decoupled GNN tensor parallelism only requires 4 times of communication.}
%For any number of model layers, decoupled GNN tensor parallelism limits the number of collective communications to 4.
%在L次NN操作开始前，\system需要进行gather操作来收集完整的顶点特征来满足保证所有NN操作的正确执行。
\subsubsection{Decoupled GNN Tensor Parallelism}
%对于用户任意的输入GNN模型，TensplitGNN将提供其对应的解耦训练模式，并应用tensor并行展开训练。具体地，在指定模型层数为L后，在每一个epoch内，\system首先对每个顶点执行L次NN操作，得到降维后的顶点嵌入。对于额外包含边相关NN操作的复杂模型，\system将进一步计算每一条超边的注意力系数，以供后续图聚合操作使用。在L次NN操作开始结束后，\system则进行split操作来恢复tensor并行，每个worker持有全部顶点的部分嵌入维度。接下来，每个worker会使用嵌入切片完成L次全图聚合操作。图聚合操作结束后，完成前向传播，\system执行gather操作来收集完整的顶点特征来满足保证LOSS函数的正确执行。反向传播则是前向传播的逆过程，\system仍然需要在L次图聚合操作前后进行嵌入的split与gather操作。

For the given input GNN model, \system provides its corresponding decoupled training mode and applies tensor parallelism for training. Specifically, after specifying the model layers $L$, in each epoch, \system first performs $L$ rounds of NN operations on each vertex to obtain the low-dimensional vertex embeddings. For complex models incorporating edge-associated NN operations, \system further computes attention coefficients for all edges to be used in subsequent graph aggregation operations. Upon completing the $L$ rounds of NN operations, \system performs a \texttt{split} operation to restore tensor parallelism, where each worker holds partial embedding dimensions for all vertices. Subsequently, each worker utilizes embedding slices to complete $L$ rounds of full-graph aggregation operations. Upon completion of the graph aggregation operations, the forward propagation is executed, followed by a \texttt{gather} operation to collect complete vertex embedding to ensure the correct execution of the loss function. Backward propagation follows the inverse process of forward propagation, where \system still needs to perform \texttt{split} and \texttt{gather} operations before and after $L$ rounds of graph aggregation operations, respectively.

%如图Figure 5所示，我们给出了在一个两层GNN模型下，朴素的GNN tensor并行与解耦的GNN tensor并行的全体通信频率对比。朴素的GNN tensor并行在两层GNN模型中需要8次全体通信（i.e., 4*L），其通信次数会随着模型层数加深而增多。相反，解耦的GNN tensor并行只需要四次全体通信，无论模型的层数是多少。另外，经过多次NN操作后，参与图聚合的顶点嵌入通常具有相比于原始特征更低的维度，进一步减少全局通信量。

% As illustrated in Figure \ref{fig:DGTP}, we provide a comparison of the overall communication frequency between naive GNN tensor parallelism and decoupled GNN tensor parallelism under a 3-layer GNN model. The naive GNN tensor parallelism requires 10 rounds of collective communication, with the communication frequency increasing as the number of model layers increases. In contrast, the decoupled GNN tensor parallelism only requires four rounds of collective communication, regardless of the number of model layers. Additionally, after multiple NN operations, the vertex embeddings involved in graph aggregation typically have lower dimensions compared to the raw features, further reducing the collective communication overhead.

% 图7是原始GNN张量并行性与解耦GNN张量并行性在整体通信频率上的比较。对于3层GNN模型，传统的GNN张量并行性需要10轮集体通信，且通信频率随模型层数增加而线性上升。相比之下，解耦的GNN张量并行性仅需四轮集体通信，且这一需求不随模型层数变化。此外，在多次神经网络操作后，参与图聚合的顶点嵌入通常具有比原始特征更低的维度，从而进一步减少了集体通信的开销。
Figure \ref{fig:DGTP} illustrates the comparison in overall communication frequency between naive GNN tensor parallelism and decoupled GNN tensor parallelism. For a 3-layer GNN model, the naive GNN tensor parallelism requires 10 rounds of collective communication, and the frequency of communication increases linearly as the number of model layers increases. In contrast, the decoupled GNN tensor parallelism only requires 4 rounds of collective communication, regardless of the number of model layers. Additionally, after multiple NN operations, the vertex embeddings involved in graph aggregation typically have lower dimensions compared to the raw features, further reducing the collective communication overhead.

\notecolor{Decoupled GNN tensor parallelism is particularly effective for message-passing based GNNs such as GCN \cite{GCN_iclr_2017}, GraphSAGE \cite{Graphsage_2017}, GAT \cite{GAT_ICLR_2018}, and GIN \cite{GNNpower_iclr_2019}. These models rely on updating and aggregating vertex features across the graph, making them ideal for GNN tensor parallelism, which efficiently partitions features and balances loads. The decoupled training method reduces communication overhead by decoupling update and aggregation processes. However, our approach may not directly benefit GNNs that do not rely on message passing, such as spectral-based GNNs (e.g., ChebNet \cite{chebnet_arxiv_2019}) and GNNs with global attention mechanisms (e.g., Graph Transformer \cite{graphtrans_arxiv_2020}). By focusing on message-passing based GNNs, NeutronTP enhances training efficiency and scalability, demonstrating broad applicability within widely-used GNN models.}

\begin{figure*}
\vspace{-0.1in}
  \centering
  \includegraphics[width=0.95\linewidth]{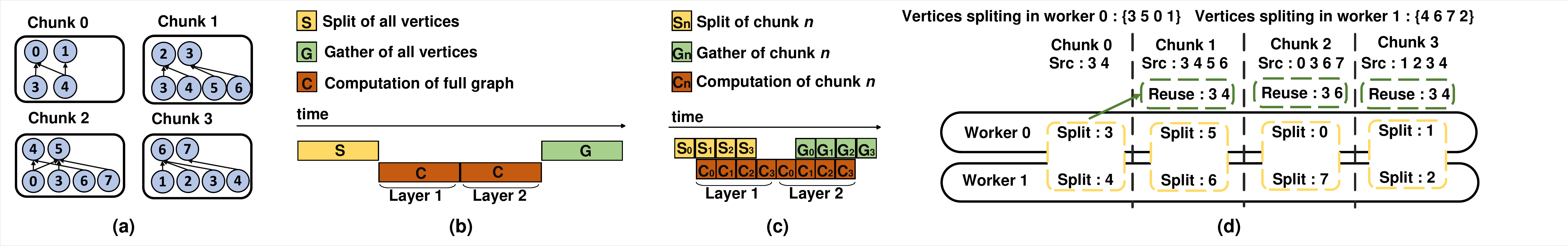}
    \vspace{-0.15in}
  \caption{(a) An example of graph data partitioned into four chunks. (b) An example of the sequential execution of different phases. (forward computation for a 2-layer GNN) (c) An example of inter-chunk pipelining. (forward computation for a 2-layer GNN) (d) An example of partitioning the \texttt{split} operation into chunk-level communication tasks.}

  \label{fig:intralayer}
  \vspace{-0.1in}
\end{figure*}

\subsubsection{Convergence Analysis}
% \textcolor{red}{to do...}

In this section, we provide a theoretical analysis of convergence guarantees for \system.
Decoupled GNN training methods have been widely used by machine learning systems \cite{simplegcn_icml_2019, sign_arxiv_2020, appnp_iclr_2019,adapro_trans_2021,DAGNN_KDD_2020}, and we present the theoretical analysis referring to the APPNP \cite{appnp_iclr_2019} and DAGNN \cite{DAGNN_KDD_2020}. 
% 现阶段的大部分下游任务可以通过输入特征来完成。在不使用任何图结构信息的情况下，仅使用NN操作的神经网络模型仍然表现良好（有一些引用就好了）。以文档分类任务为例，直观上，文档的类别完全由于其内容（即词嵌入导出的特征）来决定，而不是与其他文档的引用关系。利用邻居的特征只是平滑相似文档的嵌入，以简化文档的分类。因此，我们得到了以下假设和定理。
\notecolor{The expressive power of GNNs originates from NN operations and graph operations themselves, not their coupled execution. Thus, the decoupled GNN training method separates and sequentially executes these operations while maintaining comparable expressive power to original GNNs. Experimental results from DAGNN \cite{DAGNN_KDD_2020} demonstrate that training on the input vertex feature with NN operations can achieve a certain level of accuracy in node classification and applying graph aggregations in a decoupled manner can further enhance the performance. Therefore, we provide Assumption 1 as follows:}

%1.使用原始feature训练可以得到与原来相同的结果，MLP
\begin{itemize}[leftmargin=*]
    \item \textbf{Assumption 1} The initial features have sufficient information for the machine learning task, and the graph aggregation operation can help the model learn structural information.
    
    \item \textbf{Theorem 1} \notecolor{Under Assumption 1, the decoupled GNN training method separates graph operations from NN operations. The convergence of the decoupled GNN training can be guaranteed by the convergence of the NN and graph operations.}
\end{itemize}

% 是否需要把g_{\theta}(X)
Decoupled GNN training uses the iterative equation
\begin{align}
    \hat{L} &= \notecolor{UPDATE(X)} = MLP^{k}(X),\\
    Z^0 &= \hat{L},\\
    Z^{k} &= \notecolor{AGG(Z)} = \gamma \hat{A} Z^{k-1},
 \label{form:decoupled_itera}
\end{align}

% 代表NN Operation，即将X输入到多层的神经网络中，得到更新后的嵌入向量L，通常，这个嵌入向量L可以直接用于完成下游任务，如顶点分类、链接预测（cite）。为了学习图的结构信息，解耦GNN方法在L的基础上加入了多层的图传播操作，即AGG。
\noindent \notecolor{where $k$ represents the number of times the NN operations and graph operations are executed, $X$ is input features. $UPDATE(\cdot)$ represents NN operation, i.e., \notecolor{$X$ is fed into a multi-layer neural network to obtain the embeddings $\hat{L}$. To learn the structural information of the graph, decoupled GNN training performs multi-layer graph operations, i.e., $AGG(\cdot)$.} 
$\hat{A}$ is symmetrically normalized adjacency matrix 
($\hat{A} = \tilde{D}^{-\frac{1}{2}} \tilde{A} \tilde{D}^{-\frac{1}{2}}$, where $A$ represents the adjacency matrix, $D$ represents the degree matrix. Here, $\tilde{A}$ and $\tilde{D}$ correspond to $A + I$ and $D + I$, respectively, with $I$ being the identity matrix). 
$\gamma$ ($\gamma \in (0,1]$) is the weight of edges, which can be computed through a self-attention mechanism as in GAT or weighted neighbor convolution as in GCN.}
% % 通过邻接矩阵与边上的weight，可以表示几乎所有算法的图操作，例如GCN、GAT、GraphSAGE等。
% \notecolor{$\gamma \hat{A} Z^{k-1}$ can represent almost all graph operations in algorithms such as GCN [], GAT [], GraphSAGE [], etc

$MLP^{k}(X)$ is the convergent function. \notecolor{This is because multiple NN operations can be viewed as a traditional deep neural network model (i.e., multi-layer perceptron), whose convergence properties are well-established \cite{UAT_1998,UAT_ARXIV_2021}.} Therefore, we only need to consider $AGG(Z)$. After k iterations of graph propagation, $AGG(Z)$ can be expressed as:
\begin{align}
    Z^{k} = \gamma^k \hat{A}^k \hat{L}.
    \label{form:decouple_P}
\end{align}

If we take the limit $k \to \infty$ in Formulation \ref{form:decouple_P}, the result tends to 0 since $\gamma_k \in (0,1]$, the eigenvalues of $\hat{A}$ are the same as those of $\tilde{A}$, which can be proven through Gershgorin circle theorem \cite{weisstein2003gershgorin} that the maximum eigenvalue is 1, i.e., $\Vert \hat{A} \Vert \leq 1$, and $\hat{L}$ is convergent, resulting in
\begin{align}
    Z^{\infty} = \gamma^{\infty} \hat{A}^{\infty} \hat{L} \to 0,
\end{align}
the above concludes that the convergence is guaranteed.

In Section 5.7, we also evaluate the accuracy of decoupled GNN training, which performs comparably to coupled GNN training across different datasets.

%Several methods \cite{simplegcn_icml_2019, sign_arxiv_2020} execute graph operations in advance, and then feed the propagated features into multiple NN operations. on the contrary, several methods \cite{appnp_iclr_2019,adapro_trans_2021,DAGNN_KDD_2020} first transform the vertex features and then propagate the vertex information to distant neighbors

% \begin{figure*}
%   \centering
%   \includegraphics[width=\linewidth]{figures/intralayer.pdf}
%   \caption{(a) An example of graph data partitioned into four chunks, with edges omitted, showing only the source vertices and destination vertex sets. (b) An example of intra-layer pipelining, omitting the NN stage, showing only the overlap between graph aggregation operations and collective communication (forward propagation of a 2-layer GNN). (c) Adjusting the communication load at $t_{0}$ to split the vertices required for the next chunk in advance. (d) Adjusting the communication load at time $t_{6}$ to delay some vertex gathering tasks. }
%   \label{fig:intralayer}
% \end{figure*}

\subsection{Memory-efficient Task Scheduling Strategy}
\label{sec5}

%(a) 一个被划分为四个chunk的图数据示例，省略了边，只展示源顶点与目的顶点集合. (b) 一个intra-layer piplining的示例，省略了NN阶段，只展示图聚合操作与全体通信的重叠部分 (2-layer GNN的前向传播过程) (c) 调整t0时刻的通信负载将下一个chunk所需顶点提前split. (d) 调整t6时刻通信负载将部分顶点gather任务滞后. 

% For challenge \#2, opportunities for solutions can be found by further subdividing tasks within each worker to perform fine-grained full-graph GNN training. Therefore, we propose a memory-efficient subgraph scheduling strategy that reduces runtime memory consumption through chunk-based subgraph scheduling and further enhances performance by inter-subgraph pipeline. 

%Specifically, within each worker, we partition the entire graph into multiple chunks that can fit into GPU memory, with each chunk containing a set of vertices with contiguous IDs along with all their incoming edges. During the training process, each worker schedules chunks onto the GPU in the same order, maintaining load balance for GNN tensor parallelism. Without compromising the layer-wise synchronization barrier, the inter-layer pipeline overlaps the computation tasks of the current chunk within each layer of training with the communication tasks of the previous chunk.
% \sanzo{replot}
% \system organizes the topological structure of each subgraph chunk into compressed sparse row (CSR) format. 

\subsubsection{Chunk-based task Scheduling}
%在GNN tensor并行中，每个Worker负责部分嵌入维度的全图GNN训练。考虑到真实世界图数据庞大的尺寸，我们并不假设全图拓扑数据与顶点嵌入切片都能装入GPU设备内存。%在每个worker内,tensplitGNN应用chunk-based划分方法去把图数据切分为可以被单个GPU处理的小执行单元。每个chunk包含一部分具有连续顶点ID的顶点以及每个顶点的全部入边，便于对每一个子图进行独立的全邻居聚合。值得一提的是，我们只需要对目的地的入边进行分组，因为复杂的聚合只在前向传播中执行。在反向传播过程中，源顶点沿着出边通过求和来累积梯度。利用求和操作的关联性，不同块中的多个源顶点副本可以独立计算梯度，并随后进行求和。在GNN张量并行中使用chunk-based划分有两个优势。首先，使用chunk-based划分去获得子图是十分轻量级的，全图拓扑只需要在本地worker进行逻辑分区，而不用修改任何物理存储位置。其次，我们只需要保证每个worker以相同顺序调度chunk进行计算来保证负载均衡。我们不需要考虑chunk间的负载均衡，正如我们在section2中motivate的一样，这是一个难以达到的目标。

%此外，我们只需要保证每个worker以相同顺序调度chunk进行计算来保证负载均衡。我们不需要考虑chunk间的负载均衡，正如我们在section2中motivate的一样，这是一个难以达到的目标。

%在实际分区的过程中，不同于传统分布式训练系统在多个worker间划分子图，我们并没有指定的需要划分的子图个数。通常来说，为了更充分的利用GPU资源，减少调度开销，我们应该考虑让每个子图尽可能大。另一方面，为了构建一个有效的流水线，将GPU计算和主机-GPU通信重叠在一起 (Section 5.2)，我们最多利用 GPU 内存一半的大小来处理每个block，这样就可以在一个block被处理的同时，将另一个block载入或载出 GPU。
%这并不会造成跨worker的顶点依赖，因为每个worker只在本地以相同策略划分全图

% In GNN tensor parallelism, each worker is responsible for conducting full-graph GNN training on a portion of embedding dimensions. Considering the larger size of real-world graph data, we cannot assume that both the whole graph topology data and vertex embedding slices can fit into GPU device memory. 
% Therefore, 

% The first paragraph of Section 4.2.1: 
% Within each worker, NeutronTP employs a chunk-based partitioning approach to partition the graph data into small execution units that can be processed by a single GPU.
\notecolor{To address challenge \#2, NeutronTP employs chunk-based task scheduling, where the global graph topology is partitioned on each worker, and different workers simultaneously schedule the same subgraph. This does not incur cross-worker vertex dependencies, as each worker partitions the entire graph locally using the same strategy. All workers execute communication and computation tasks for each chunk in the same order to ensure load balancing. }
As shown in Figure \ref{fig:intralayer} (a), each chunk comprises a subset of destination vertices with contiguous vertex IDs and all in-edges for each vertex, facilitating independent full-neighbor aggregation for each chunk. 
In GNN tensor parallelism, using chunk-based partitioning offers two advantages. Firstly, using chunk-based partitioning to obtain subgraphs is lightweight, as the graph topology only needs to be logically partitioned within local workers without modifying any physical storage locations. 
% Secondly, ensuring load balance merely requires scheduling chunks for computation in the same sequence across all workers. 
% 其次，确保负载平衡只需要在所有工作线程中以相同的顺序调度计算块，从而消除了处理块之间的负载平衡的需要（第 2.2 节）。
Secondly, ensuring load balancing merely requires scheduling chunks for computation in the same order across all workers, thus eliminating the need to handle load balancing between chunks.
% Load balancing between chunks does not need to be considered, as motivated in Section 2.2, which remains a challenging goal to achieve. 

It is worth mentioning that we only need to group the in-edges of destinations, as complex aggregations are performed only during the forward pass. During the backward pass, source vertices can accumulate gradients along out-edges through summation. Leveraging the associativity of summation operations, multiple copies of source vertices in different chunks can independently compute gradients and then be summed afterward. During the actual partitioning process, unlike traditional distributed training systems where chunks are divided among multiple workers, we do not specify a predetermined number of chunks to partition. Generally, to better utilize GPU resources and reduce scheduling overhead, we should aim to make each chunk as large as possible. 

%另外，在支持复杂模型时，我们需要将历史嵌入保存在CPU去计算下一轮的注意力系数。将NN操作下推至CPU执行也可以便利这一过程，因为我们不需要额外传输嵌入回到CPU。

%当以chunk为调度单位时，full-graph GNN训练还需要考虑中间结果的管理。中间结果在前向计算中产生并在反向传播的梯度计算中使用。为了避免中间结果超过GPU内存容量，我们需要将其传回CPU，并在反向计算时加载到GPU使用。这种频繁的host-GPU数据交换可能会影响总体性能。幸运的是，受益于解耦GNN训练方式，\system中连续多次图聚合操作不产生中间结果，我们只需要处理NN阶段产生的中间结果。考虑到GNN训练的计算开销通常在图聚合操作而不是NN操作，我们将NN相关操作下推至CPU执行。这不仅节省了大量中间结果的host-GPU通信，还利用了CPU资源。

When using chunks as the scheduling unit, full-graph GNN training requires consideration of intermediate result management. Intermediate results are generated during forward computation and consumed in the gradient computation during backward pass \cite{hongtu_SIGMOD_2024}. To avoid exceeding the GPU memory capacity with intermediate results, they need to be sent back to CPU memory after forward computation and returned to GPU during backward computation. This frequent host-GPU data exchange may impact overall performance. Fortunately, benefiting from the decoupled GNN training approach, \system avoids generating intermediate results during consecutive graph aggregation operations. Thus, we only need to handle intermediate results produced during the NN operations. Considering that the computational overhead of GNN training typically lies in graph aggregation operations rather than NN operations \cite{exp_vldb_2024}, we push down NN operations to be executed on the CPU. This not only reduces a significant amount of host-GPU communication for intermediate results but also leverages CPU resources.

% Additionally, when supporting complex models, we need to retain historical embeddings on the CPU to compute the attention coefficients for the next round. Pushing NN operations to be executed on the CPU facilitates this process as we do not need to transfer embeddings back to the CPU separately.

% On the other hand, to construct an effective pipeline that overlaps GPU computation with host-GPU communication, we utilize at most half the size of GPU memory to handle each chunk. This allows for loading or unloading another chunk onto the GPU while one chunk is being processed.

%
%在chunk-based子图调度的基础上，我们可以进一步重叠每一个chunk的通信与计算过程。我们计划将每一个split或gather操作与相邻的一层图聚合操作重叠，并且不破坏原本的layer-wise同步。如图8(b)所示，split操作会提前为每个chunk切分src顶点的嵌入，而gather操作在chunk计算结束后为每个chunk收集dst顶点的嵌入切片。然而，这种chunk级别的通信可能会破坏原本的负载均衡。在开始训练前，我们为每个worker分配顶点相关的通信任务。如图8(c)所示，worker0负责顶点0，2和4，而worker1负责顶点1，3和5。当以chunk为单位执行split与gather操作时，我们无法确定src与dst顶点在不同worker的分布情况，造成通信负载不均。例如，当为chunk 0执行split任务时，worker 0内的切分顶点有两个（i.e, 0 and 4），而worker 1内的切分顶点只有一个(i.e.,1)。

%我们采用了一个两级的通信调整策略来确保chunk级别的通信负载均衡。首先，为了避免chunk内连续ID顶点的通信任务集中在某些worker内，我们采用一个hash策略去均匀分配顶点相关通信任务到不同的worker。其次，我们进一步调整每次split与gather操作所涉及的顶点集合，保证每个worker内的顶点相关通信任务均衡。如图Figure 6（c）所示，在split操作中，我们会在一些拥有更少通信任务的worker(i.e, worker 1)内提前split下个chunk所需节点。具体的，我们提前在t0时刻提前为chunk1完成了顶点3的split任务。在gather操作中，我们将一些拥有更多通信任务的worker(i.e, worker 0)内的gather任务滞后，如图Figure 6 (d)所示。具体的，我们在t6时刻将chunk0的顶点2的gather任务滞后到t7时刻。注意这并不会影响整体计算过程，因为gather操作是为后续的NN操作准备输入数据。对于任意输入图数据，我们只需要调整通信负载一次，并在每次迭代中直接使用已经调整好的通信负载。基于上述通信调整策略，我们将split与gather过程拆分并与不同子图的图聚合操作重叠，如图Figure 6 (b) 所示。

\subsubsection{Inter-chunk Pipelining}
%在chunk-based子图调度的基础上，我们可以进一步重叠每一个子图的通信与计算过程。我们计划将每一个split或gather操作与相邻的一层图聚合操作重叠，并且不破坏原本的layer-wise同步。具体地，我们以Figure 5 (b)为例，图中的split操作将与其右侧的图聚合操作重叠，而gather操作将与其左侧的图聚合操作重叠。为此，我们需要将split与gather操作都划分为与子图计算任务相对应的通信任务，满足每个子图计算所需的输入数据。然而，如果只是简单的按照子图划分通信任务，则可能造成通信负载不均衡的问题。如我们在Section 3.2所讨论的，在以全图为通信单位时，每个worker每次进行split或gather操作的通信量可以表示为V/W * M/W *(W-1),其中V表示顶点总数，M表述特征维度，W表示worker数。当以子图为通信单位时，则会改变每个worker通信量中的V/W，并取决于每个子图所涉及到的源顶点或目的顶点集和。如图Figure 6 (c) 与 (d) 所示，chunk 0相关的split与gather任务在两个worker上的通信负载并不均衡。在图Figure 6 (b)的t0时刻，每个worker需要进行Split操作为chunk0的源顶点集合{0 1 4}准备特征切片。在图Figure 6 (b)的t6时刻，每个worker需要进行gather操作连接目的顶点集合{0 1 2}的嵌入切片，作为后续NN操作的输入。在实际训练中，这个过程会变得更复杂

%在chunk-based子图调度的基础上，我们可以进一步重叠每一个chunk的通信与计算过程。我们计划将每一个split或gather操作与相邻的一层图聚合操作重叠，并且不破坏原本的layer-wise同步。如图8(b)所示，全图计算过程与两个全体通信操作需要串行执行来保证layer-wise的同步。受益于chunk-based的任务调度，我们可以进一步将两个全体通信操作划分为chunk级别的通信任务来与计算任务重叠。如图8(c)所示，split操作提前为每个chunk切分src顶点的嵌入，而gather操作在chunk计算结束后为每个chunk收集dst顶点的嵌入切片。这种chunk级别的通信任务还需要进一步设计来避免冗余通信。首先，对于每个chunk，我们会将其顶点相关通信任务均分到全部worker，确保通信负载均衡。然而，由于chunk内的src顶点集合有重复，这会导致对某些顶点的嵌入切片重复进行通信操作。因此，在分配每个chunk的通信任务时，\system会检查当前chunk内顶点是否包含在之前的chunk中。如图8(d)所示，当检查到顶点已经被通信完毕，\system会直接重用之前的通信结果。另外，在遍历一遍chunk并确定每个worker内负责的顶点相关通信任务后，我们会将其打包为一个不重复的顶点集合。每个worker会根据本地的顶点集合执行相关通信任务与NN计算任务。

\begin{algorithm}[t]\small
	\caption{Workflow of \system for a single epoch}  
	\label{alg:exec_flow}  
	\begin{algorithmic}[1]
		\Require Graph $G(V,E)$, Feature $\textbf{h}^{0}$,
		Initial parameterized GNN layers $\{\textbf{W}^0, \textbf{W}^1 \ldots \textbf{W}^{L-1}\}$, cluster size $m$,  chunk number $n$
		\Ensure Updated parameterized GNN layers $\{\textbf{W}^0, \textbf{W}^1 \ldots \textbf{W}^{L-1}\}$
        \State $\{G_{j}$$\mid$$ 0$$\leq$$j$$<$$n\}$=\textbf{\texttt{chunk\_partition}}($G$, $n$)
        \State $\{V_{i}, \textbf{h}_{i}^{0}$$\mid$$ 0$$\leq$$i$$<$$m\}$=\textbf{\texttt{distribute\_vertex}}($\{G_{j}$$\mid$$ 0$$\leq$$j$$<$$n\}$, $\textbf{h}^0$, $m$)
        \For{\texttt{worker} $i=0$ to $m-1$} \textbf{in parallel}
	    \For{layer $l$ = $0$ to $L-1$}  
        \State {$\textbf{h}^{l+1}$  of $V_{i}$ =\enspace \texttt{worker(}$i$\texttt{).UPDATE}($\textbf{W}^l$, $\textbf{h}^{l}$ of $V_{i}$)}
        \EndFor
        \For{layer $l$ = $0$ to $L-1$}  
        \For{\emph{chunk} with id $j$ = $0$ to $n-1$}
        \If{layer == $0$}
        \State $\textbf{h\_cut}^{0}_{i}$ of ${N_{j}}$$\leftarrow$\textbf{\texttt{Split}}($\textbf{h}^{L}$ of ${N_{j}}$)
        \EndIf
        \State {$\textbf{h\_cut}_{i}^{l+1}$ of $V_{j}$ =\enspace \texttt{worker(}$i$\texttt{).AGG}($\textbf{h\_cut}_{i}^{l}$ of ${N_{j}}$, ${G_{j}}$)}
        \If{layer == $L-1$}
        \State $\textbf{h}^{L}$ of $V_{j}$$\leftarrow$\textbf{\texttt{Gather}}($\textbf{h\_cut}^{L}_{i}$ of $V_{j}$)
        \EndIf
	  \EndFor
        \EndFor
		\EndFor
		\State $\textbf{loss}$ = \texttt{downstream\_task}($\textbf{h}^{L}$)
		\State $\nabla \textbf{h}^{L}$ = $\textbf{loss}$\texttt{.backward()} 
		\For{\texttt{worker} $i=0$ to $m-1$} \textbf{in parallel}
	  \For{layer $l$ = $L-1$ to $0$}  
        \For{\emph{chunk} with id $j$ = $0$ to $n-1$}
        \If{layer == $L-1$}
        \State $\nabla \textbf{h\_cut}^{0}_{i}$ of ${V_{j}}$$\leftarrow$\textbf{\texttt{Split}}($\nabla \textbf{h}^{L}$ of ${V_{j}}$)
        \EndIf
        \State {$\nabla \textbf{h\_cut}_{i}^{l}$ of $N_{j}$ =\enspace \texttt{worker(}$i$\texttt{).AGG}($\nabla \textbf{h\_cut}_{i}^{l+1}$ of ${V_{j}}$, ${G_{j}}$)}
        \If{layer == $0$}
        \State $\nabla \textbf{h}^{L}$ of $N_{j}$$\leftarrow$\textbf{\texttt{Gather}}($\nabla \textbf{h\_cut}^{L}_{i}$ of $N_{j}$)
        \EndIf
	  \EndFor
        \EndFor
        \For{layer $l$ = $L-1$ to $0$}  
        \State {$\nabla \textbf{h}^{l+1}$  of $V_{i}$ =\enspace \texttt{worker(}$i$\texttt{).UPDATE}($\textbf{W}^l$,$\nabla \textbf{h}^{l}$  of $V_{i}$)}
        \EndFor
		\EndFor
    \For{layer $l$ = $0$ to $L-1$}
        \State \texttt{sync\_and\_update} ($\textbf{W}^l$) \textcolor{gray}{//parameter update}
    \EndFor
	\end{algorithmic} 
\end{algorithm}

Based on chunk-based task scheduling, we can further overlap the communication and computation processes of each chunk. Our plan involves overlapping each \texttt{split} and \texttt{gather} operation with the adjacent graph aggregation operation without disrupting the original layer-wise synchronization. As illustrated in Figure \ref{fig:intralayer} (b), the full graph computation process and two collective communication operations need to be executed serially to ensure layer-wise synchronization. Benefiting from chunk-based task scheduling, we can further partition the two collective communication operations into chunk-level communication tasks to overlap with computation tasks. As shown in Figure \ref{fig:intralayer} (c), the \texttt{split} operation pre-splits the embeddings of src vertices for each chunk, while the \texttt{gather} operation collects the embedding slices of dst vertices for each chunk. This chunk-level communication task requires further design to avoid redundant communication. Firstly, for each chunk, we evenly distribute its vertex-related communication tasks across all workers to ensure communication load balancing. However, due to duplicate src vertex sets within chunks, this may result in redundant communication operations for some vertices. Therefore, when assigning communication tasks for each chunk, \system checks whether the vertices within the current chunk have already been communicated in previous chunks. As illustrated in Figure \ref{fig:intralayer} (d), upon detecting that the vertices have been communicated, \system directly reuses the previous communication results. After traversing each chunk and determining the vertex-related communication tasks assigned to each worker, we aggregate them into a non-redundant vertex set. Each worker executes the relevant communication tasks and NN computation tasks based on its local vertex set.

\subsection{Overall Execution Flow in \system}
%算法1概述了整个执行流程。在开始前，\system采用chunk-based任务调度策略将图拓扑逻辑划分为一系列子图，包含一系列不相交的顶点序列和他们的全部入边。接下来，\system会把chunk内的顶点相关任务依次均分到全部worker。在去重后，每个worker得到其本地顶点工作队列，用于后续执行NN计算与通信操作。在前向过程中，每个worker会首先完成L次NN操作得到本地节点的嵌入表示。接下来，全体worker按相同顺序调度chunk，并首先切分其入邻居集合的嵌入。切分结束后，每个worker使用本地嵌入切片开始图聚合计算。在chunk的L次图聚合结束后，全体worker会开始gather操作，收集其dst顶点的完整嵌入。在完成全部chunk的计算与通信任务后，每个worker开始下游任务并计算梯度。反向过程是前向过程的逆过程，同样需要在图聚合操作前后进行嵌入切分与聚合操作。
Algorithm 1 outlines the overall execution flow. To begin with, \system employs a chunk-based task scheduling strategy to partition the graph topology into a series of chunks $G_j$, each containing a set of disjoint vertex sets $V_{j}$ and their in-neighbor sets $N_{j}$, where $j$ is the chunk id (line 1). 
Subsequently, \system evenly distributes the vertex-related tasks within each chunk across all workers. After deduplication, each worker obtains its local vertex work queue $V_{i}$, where $i$ is the worker id, utilized for subsequent execution of NN computations and communication operations (line 2). 
During the forward pass, each worker first completes $L$ rounds of NN operations to obtain local vertex embeddings (line 5). Next, all workers schedule chunks in the same order and begin by splitting the embeddings of their in-neighbor sets, i.e., $\textbf{h}^{L}$ of $N_{j}$ (Line 9).
Upon completion of the splitting phase, each worker initiates graph aggregation operations using its local embedding slices, i.e., $\textbf{h\_cut}_{i}^{l}$ of ${N_{j}}$ (Line 10). 
Following $L$ rounds of graph aggregation within the chunk, all workers gather the complete embeddings of destination vertices in each chunk, i.e., $\textbf{h}^{L}$ of $V_{j}$ (Line 12). 
Upon completion of all chunk computations and communication tasks, each worker initiates downstream tasks and computes gradients $\nabla \textbf{h}^{L}$ (lines 13-14).
The backward pass is the reverse process of the forward pass, requiring embedding \texttt{split} and \texttt{gather} operations before and after the graph aggregation operation, respectively.

\section{Evaluation}
\label{sec5}

\subsection{Experimental Setup}
\vspace{-0.05in}

\Paragraph{Environments.} Our experiments are conducted on the Aliyun ECS cluster with 16 GPU nodes. Each node (ecs.gn6i-c16g1.4xlarge instance) is equipped with 16 vCPUs, 186GB DRAM, and 1 NVIDIA Tesla T4 GPU, running Ubuntu 18.04 LTS OS. The network bandwidth is 15 Gbps/s. Libraries CUDA 11.1, OpenMPI-3.0.2, PyTorch v1.5 backend, and cuDNN 7.0 are used in both clusters.

%softeware of \sysname

\Paragraph{Datasets and GNN algorithms.}
Table \ref{tab:Dataset} lists six graph datasets that we used in our evaluation, including three popular GNN datasets: Reddit \cite{Graphsage_2017}, Ogbn-products \cite{ogb_2020_neurIPS}, and Ogbn-paper \cite{ogb_2020_neurIPS}, and one graph dataset Friendster \cite{FS_ARXIV_2016}. Ogbn-mag \cite{ogb_2020_neurIPS} and Mag-lsc \cite{ogblsc_nips_2021} are two heterogeneous graphs that we use to evaluate the heterogeneous GNN training. For graphs without ground-truth properties (Friendster), we use randomly generated features, labels, training (65\%), test (10\%), and validation (25\%) set division.
We use two popular GNN models with different computation patterns, GCN \cite{GCN_iclr_2017} and GAT \cite{GAT_ICLR_2018}. All of them are in a 2-layer structure. The vertex feature dimensions, hidden layer dimensions, and the number of labels of datasets are listed in Table \ref{tab:Dataset}.

\Paragraph{Competitor systems.}
In our performance evaluation, we compare \system with two kinds of GNN training systems, i.e., mini-batch system and full-graph system.
For the mini-batch system, we compare \system with DistDGL \cite{distdgl_sc_2020}, a representative deep learning library for graphs. DistDGL relies on data sampling to reduce computation cost \cite{distdgl_sc_2020}, which is set to execute a (25, 10) neighborhood sampling for the training. In such a configuration, DistDGL picks a maximum of 10 neighbors for the first hop of a vertex, and then a maximum of 25 neighbors for each of those 10. 
For the full-graph system, we compare \system with NeutronStar \cite{neutron_sigmod_2022} and Sancus \cite{SANCUS_VLDB_2022}. Neutronstar \cite{neutron_sigmod_2022} designs hybrid vertex dependency management for a more balanced use of communication and computational resources. Sancus \cite{SANCUS_VLDB_2022} reuses historical embedding for cross-worker vertices to reduce communication.

\begin{table}[!t]
\vspace{-0.1in}
	\caption{Dataset description.}
	\vspace{-0.1in}
	%\hspace{-0.5in}
	\label{tab:Dataset}
	\centering
	\small
    \footnotesize
	{\renewcommand{\arraystretch}{1.2}
	\begin{tabular}{l r r c c c c c}
		\hline
		
		\hline
		{\textbf{Dataset}} &
		{\textbf{|V|}} &
		{\textbf{|E|}}  &
		{\textbf{ftr. dim}}&
		{\textbf{\#$\mathbb{L}$}}&
        {\textbf{hid. dim}}\\
		\hline
%		\hline
		{Reddit (RDT)} & 0.23M & 114M &602&41 &256\\		
%		\hline
		{Ogbn-products (OPT)}  &2.45M& 61.68M &100&47&64\\
%		\hline
		{Ogbn-paper (OPR)}& 111.1M &1.616B&128&172 &128\\
%		\hline
		{Friendster (FS)}& 65.6M &2.5B&256&64 &128\\
          \notecolor{{Ogbn-mag (MAG)}}& \notecolor{1.9M} &\notecolor{21M}&\notecolor{128}&\notecolor{349} &\notecolor{64}\\
        \notecolor{Mag-lsc (LSC)}& \notecolor{244.2M} &\notecolor{1.7B}&\notecolor{768}&\notecolor{153} &\notecolor{256}\\
		\hline
		
		\hline
		
	\end{tabular}
	}
	\vspace{-0.1in}
\end{table}

\subsection{Overall Comparison}
%我们与neutronstar，distDGL与sancus在16个节点的集群中对比了总体性能。我们使用了GCN与GraphSAGE作为训练模型，并记录计算时间，通信时间与每个epoch的总运行时间.‘comp. max’ 与‘comm. max’分别代表全部worker中最长计算时间与通信时间,通常决定着分布式训练实际的运行时间。同样，"comp.min "和 "comm.min "表示所有工作者中计算时间和通信时间最短的。对于采用overlap技术重叠计算与通信的系统（i.e., NeutronStar and NeutronTP）,他们的‘comp. max’ 与‘comm. max’之和会大于总运行时间。通过仔细记录这些时间，我们可以深入了解分布式训练的工作量。实验结果汇总如表所示。

We conduct a comprehensive performance comparison with NeutronStar \cite{neutron_sigmod_2022}, DistDGL \cite{distdgl_sc_2020}, and Sancus \cite{SANCUS_VLDB_2022} on a 16-node cluster. We record the computation time, communication time, and the per epoch runtime. 
"max" indicates the longest computation and communication time among all workers, which typically determines the actual runtime of distributed training. Similarly, "min" indicates the shortest computation time and communication time among all workers. For systems employing pipelining techniques to overlap computation and communication (i.e., NeutronStar \cite{neutron_sigmod_2022} and NeutronTP), the sum of the longest computation and communication time exceeds the total runtime. By meticulously logging these times, we gain insights into the workload of distributed training. The experimental results are summarized in Table \ref{tab:overall}.

%相比于DistDGL，TenparaGNN在大多数数据集中表现得更好，并获得了最高12.2倍的加速比。DistDGL使用mini-batch训练方式，每个batch内全部worker会对本地节点采样获得子图并在子图上进行训练。在Secition 2中我们讨论过顶点依赖会随着图分区数量增多而上升，而不同batch内的采样子图等同于将图数据进一步划分而造成更多的重复计算与通信。另一方面，mini-batch训练无法保证计算负载均衡因为不同worker内采样子图规模并不相同。同时，DistDGL使用的metis划分可能使某些worker具有更多顶点，从而在特征通信过程被其他worker访问更多次。如Table 2所示，DistDGL在不同worker间的计算时间差距与通信时间差距分别可以达到最大7.1倍与2.2倍，造成负载较低worker内的资源浪费。\system通过采用tensor parallelism的分布式训练方式，获得更均衡的工作负载同时避免冗余计算与通信。在 Ogbn-paper 上，DistDGL 实现了更好的性能，因为它只使用了 120 万个顶点（1.1%）进行训练，与 \system 相比，计算量大大降低。但如果我们从吞吐量的角度考虑，即单位时间处理的顶点数，\system仍然具有显著优势。当需要从全图范围学习全局特征和模式，高吞吐量意味着更快的处理速度。

\Paragraph{Comparison with mini-batch system.} 
Compared to DistDGL \cite{distdgl_sc_2020}, \system exhibits superior performance across most datasets, achieving up to 6.23$\times$ speedup. 
% DistDGL adopts a mini-batch training approach, where each batch involves all workers sampling local vertices to obtain subgraphs and conducting training on these subgraphs. 
% As discussed in Section 2.2, the vertex dependencies increase with the number of graph partitions, and the sampled subgraphs within different batches further partition the graph, resulting in more redundant computations and communications. 
The METIS partitioning used by DistDGL may result in certain workers having more vertices, leading to more frequent access by other workers. As shown in Table \ref{tab:overall}, DistDGL exhibits an imbalance in both computation and communication times among different workers, with disparities reaching up to 1.38$\times$ and 2.2$\times$, respectively, thereby causing resource wastage in the less loaded workers. \system mitigates these issues by adopting tensor parallelism, achieving more balanced workloads while avoiding redundant computations and communications. On the Ogbn-paper dataset, DistDGL demonstrates a better performance as it trains only on 1.1\% of the total vertices, resulting in reduced computational load compared to \system. 
% However, from a throughput perspective, i.e., the number of vertices processed per second, \system still maintains a significant advantage (i.e., average 650K vertices/s vs 50K vertices/s). When learning global patterns across the entire graph is required, higher throughput implies faster processing speeds.

\begin{table}[t] 
\vspace{-0.1in}
	\caption{Comparison with different systems a 16-node ECS cluster. "max" and "min" indicate the longest and shortest computation or communication times among workers, respectively. "total" indicates the per-epoch runtime. }
 %\zyf{why not show accuracy? what not use fig?}
	\label{tab:overall}
	\centering
	\footnotesize
	{\renewcommand{\arraystretch}{1.0}
    \resizebox{\linewidth}{!}{\begin{tabular}{c|c|c|c c c c l}
		\hline
		
		\hline
		\multirow{3}*{Model}&\multirow{3}*{ Dataset}&\multirow{3}*{System}
		&\multicolumn{5}{c}{ Runtime (s)} \\
		\cline{4-8}
        &&&\multicolumn{2}{c}{Computation}&\multicolumn{2}{c}{Communication} &\multirow{2}*{\textbf{total}}\\
		\cline{4-7}
		&&&{\textbf{max}} &
		{\textbf{min}}  &
        {\textbf{max}}&
		{\textbf{min}} \\
		\hline

        \hline
        \multirow{16}*{GCN}&
        \multirow{4}*{RDT}
		&DistDGL &0.15 &0.11 &2.12 & 1.38 & 2.27\\
        &&NeutronStar&0.86 &0.77 &1.17 & 0.87 & 1.92\\
        &&Sancus  &0.35 &0.31 &0.82 & 0.71&1.17\\
        \cline{3-8}
        &&\system &0.39 & 0.38& 0.19&0.18&\textbf{0.40}\\
		\cline{2-8}
    
        \cline{2-8}
        &\multirow{4}*{OPT}
		&DistDGL &0.26 & 0.16 &2.82 & 1.28 &3.18\\
        &&NeutronStar&2.71 & 1.42 &2.89 &1.78 &4.45\\
        &&Sancus  &0.86 & 0.36 &1.59 &1.22 &2.45\\
        \cline{3-8}
        &&\system &0.46 & 0.44 &0.24 &0.22 &\textbf{0.50}\\
        \cline{2-8}
    
        \cline{2-8}
        &\multirow{4}*{OPR}
		&DistDGL &5.35 &4.19&20.1 &11.21&\textbf{25.4}\\
         &&NeutronStar&-&-&-&-&OOM\\
        &&Sancus  &-&-&-&-&OOM\\
        \cline{3-8}
        &&\system &95.8 & 95.2& 53.6& 49.4 &134.4\\
        \cline{2-8}
    
        \cline{2-8}
        &\multirow{4}*{FS}
		&DistDGL &136.4 &118.9&323.4&197.5&459.5\\
         &&NeutronStar&-&-&-&-&OOM\\
        &&Sancus  &-&-&-&-&OOM\\
        \cline{3-8}
        &&\system &74.3&73.5 & 32.9&29.4 &\textbf{90.5}\\
        \hline
    
		\hline

        \multirow{16}*{GAT}&
        \multirow{4}*{RDT}
		&DistDGL &0.75 &0.52 &2.17 & 1.49 & 2.92\\
        &&NeutronStar&- &- &- & - & OOM\\
        &&Sancus  &-&-&-&-&OOM\\
        \cline{3-8}
        &&\system &0.92 & 0.88& 0.48&0.42&\textbf{1.29}\\
		\cline{2-8}
    
        \cline{2-8}
        &\multirow{4}*{OPT}
		&DistDGL &1.17 & 0.94 &2.76 & 1.29 &3.93\\
        &&NeutronStar&8.72 & 5.98 &15.9 &8.29 &22.4\\
        &&Sancus  &-&-&-&-&OOM\\
        \cline{3-8}
        &&\system &2.17 & 1.94 &1.06 &0.95 &\textbf{3.03}\\
        \cline{2-8}
    
        \cline{2-8}
        &\multirow{4}*{OPR}
		&DistDGL &8.40 &6.48& 21.1&11.7&\textbf{29.5}\\
         &&NeutronStar&-&-&-&-&OOM\\
        &&Sancus  &-&-&-&-&OOM\\
        \cline{3-8}
        &&\system &154.3 & 136.4& 98.9& 84.7 &235.4\\
        \cline{2-8}
    
        \cline{2-8}
        &\multirow{4}*{FS}
		&DistDGL &157.8 &110.4&419.8 &283.7&577.6\\
         &&NeutronStar&-&-&-&-&OOM\\
        &&Sancus  &-&-&-&-&OOM\\
        \cline{3-8}
        &&\system &115.2 & 92.5 &72.1 & 61.4 & \textbf{167.9}\\
        \hline
    
		\hline
	\end{tabular}
 }}
\end{table}

%相比于NeutronStar和Sancus，\system在所有数据集中都表现更好，并分别取得了最大8.9倍与4.9倍的加速比。他们都受限于数据并行造成的不平衡工作负载和大量跨worker的顶点依赖造成的通信开销。NeutronStar和Sancus在不同worker间的计算时间差距分别可以达到最大7.1倍与2.2倍，通信时间差距分别可以达到最大7.1倍与2.2倍. NeutronStar具有较长的通信时间因为它使用了chunk-based划分策略，该策略具有较多的跨worker顶点依赖。尽管NeutronStar使用ring-based的通信调度策略较好的重叠了计算与通信时间，但较高的通信开销仍然限制他的性能。Sancus通过重用历史嵌入减少了通信开销。然而，在更新历史嵌入时，Sancus会顺序启动每个worker进行嵌入广播，将本地全部嵌入发给全部worker，不管其他分区是否包含这些顶点。这不仅造成worker间长时间的相互等待，还导致许多冗余通信。另外，NeutronStar与Sancus在处理大规模图时都面临着out-of-memory错误因为每个worker内全部子图的大小很容易超过GPU内存。\system的优势来自于以下两点：（1）tensor parallelism训练获得了更均衡的计算与通信负载，同时解耦的GNN训练方式显著降低了张量并行的通信开销（2）子图调度策略将本地图数据分为多个小于GPU内存的chunk并顺序加载，重叠计算与通信的同时避免out-of-memory。 虽然这种内存交换会降低吞吐量，但我们认为在内存限制和训练效率之间进行权衡是必要的。
\Paragraph{Comparison with full-graph system.} 
Compared to NeutronStar \cite{neutron_sigmod_2022} and Sancus \cite{SANCUS_VLDB_2022}, \system demonstrates superior performance across all datasets, achieving speedups of up to 8.72$\times$ and 4.81$\times$, respectively. They are both constrained by imbalanced workloads and communication overhead resulting from extensive cross-worker vertex dependencies.
The computation and communication time gap between different workers in NeutronStar can reach up to 1.91$\times$ and 1.62$\times$, respectively. For Sancus, these gaps can reach up to 2.38$\times$ and 1.18$\times$, respectively. 
NeutronStar exhibits long communication times due to its chunk-based partitioning strategy, which has more cross-worker vertex dependencies, as described in Section 2.2. 
% Despite employing a ring-based communication scheduling strategy that effectively overlaps computation and communication, the higher communication overhead still constrains its performance. 
Sancus reduces communication overhead by reusing historical embeddings. However, when updating historical embeddings, Sancus sequentially triggers each worker to broadcast embeddings, sending all local embeddings to all workers, regardless of whether other partitions contain these vertices. This not only leads to prolonged waiting times for workers but also results in considerable redundant communication. Both NeutronStar and Sancus encounter out-of-memory errors when dealing with large-scale graphs and complex model due to their lack of intra-worker task scheduling strategies.
The advantages of \system stem from two main factors: (1) Tensor parallelism training achieves a more balanced computation and communication load, while the decoupled GNN training approach significantly reduces communication overhead. (2) The chunk-based task scheduling strategy partitions local graph data into multiple chunks smaller than GPU memory and loads them sequentially, overlapping computation and communication while avoiding out-of-memory errors. 
% Although this memory swapping may reduce performance, we believe that striking a balance between memory constraints and training efficiency is necessary.

% 性能实验
% 实验目的：验证系统性能
% 对比系统：TenparaGNN、NeutronStar、Sancus、DistDGL 
% 数据集：reddit、products、papers100M、friendster
% 实验配置：16节点，T4 GPU
% 模型：GCN、GAT
% 实验内容：对比不同系统的单个epoch运行时间

\begin{figure}
\vspace{-0.1in}
  \centering
  \includegraphics[width=\linewidth]{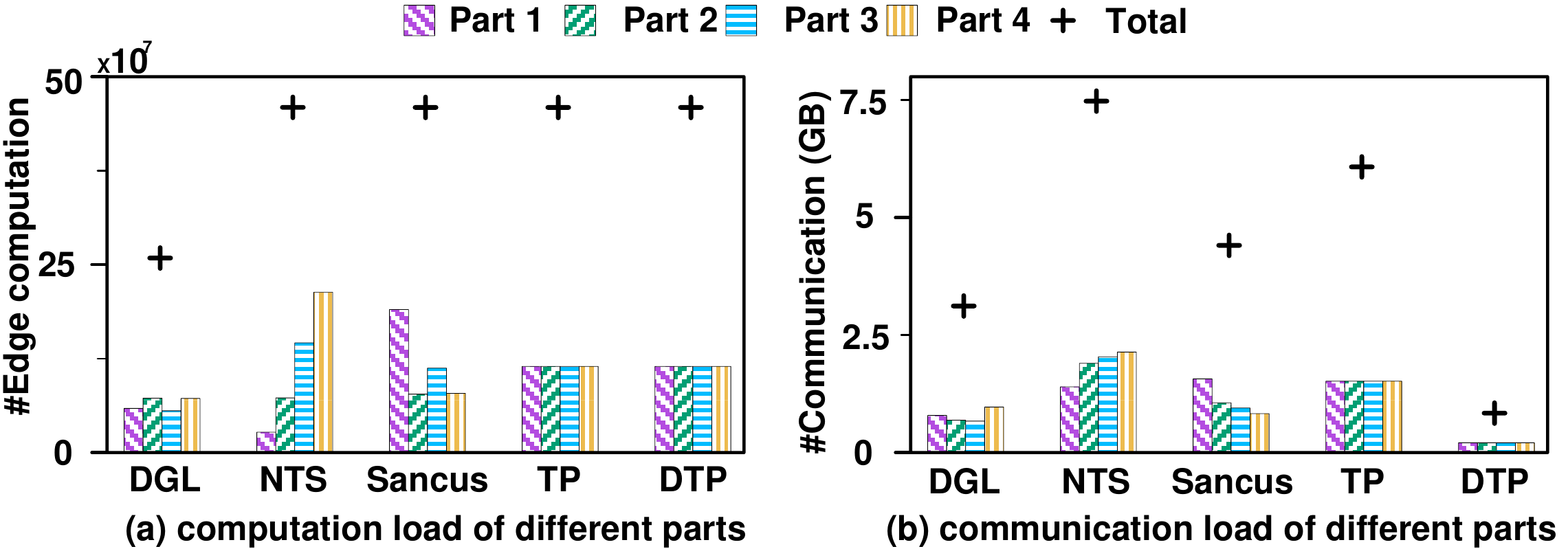}
  \vspace{-0.2in}
  \caption{The computation/communication load of each partition in different systems. "TP" indicates \system with naive GNN tensor parallelism and "DTP" indicates \system with decoupled GNN tensor parallelism.}
  \label{fig:workload-compara}
  \vspace{-0.2in}
\end{figure}

% \zyf{The computation/communication load of each partition in different systems. what are TP and DTP?}

%我们统计了不同系统的计算负载与通信负载来说明\system在负载均衡方面的优势。我们使用DistDGL，NeutronStar（NTS），Sancus，使用朴素张量并行的\system （TP），使用解耦张量并行的\system (DTP)在Reddit数据集上进行实验。为了方便观察，我们在4节点集群中训练2-layer GCN模型，并记录每个worker的工作负载与总负载。通信负载根据实际传输量统计。计算负载根据参与计算的边数统计。因为\system在计算过程中使用feature切片，而其他系统使用完整特征，我们会根据相应比例缩放\system的边计算量。实验结果如图7所示。在计算负载方面，\system通过划分规则的特征而不是复杂的图结构实现了完全的计算负载均衡。在总计算量方面，\system与其他full-graph系统持平，但高于mini-batch系统DistDGL。DistDGL通过采样降低了总计算量，但采样过程同样有不可忽视的开销，并可能与训练过程争用系统资源。在通信负载方面，\system通过保证每个worker负责相同顶点数量的特征聚集与切分任务来确保负载均衡。然而朴素的张量并行需要在每一层都执行特征的聚集与切分操作，这带来显著的通信开销，并可能超过数据并行方法的通信量。如图Figure 3（b）朴素的张量并行总通信量大于基于data parpllelism的Sancus。为此，我们采用解耦张量并行训练方法，减少全体通信次数并将通信对象降维成更轻量的顶点嵌入，显著减少通信量达7.1倍

%将NN操作与图操作分离，并将全体通信次数限制到4次，显著降低了通信量与通信频率。另外，连续的NN操作通常将原始feature降维成更轻量的顶点嵌入，这帮助DTP进一步降低通信量。

%得益于解耦张量并行化训练方法，system降低了整体通信频率同时将通信对象降维成更轻量的顶点嵌入，显著减少通信量达7.23倍。
\subsection{Computation and Communication Analysis}

We evaluate the computational and communication loads of different systems to illustrate the load-balanced advantages of \system. Experiments are performed on the Reddit dataset using DistDGL, NeutronStar (NTS), Sancus, \system with naive GNN tensor parallelism (TP), and \system with decoupled GNN tensor parallelism (DTP). We train a 2-layer GCN on a 4-node cluster and record the workload for each worker as well as the overall workload. The communication load was measured based on the amount of data transferred, while the computational load was determined by the number of edges involved in the computation. Since \system utilizes feature slices during computation, whereas other systems use complete features, we scale the edge computation of \system accordingly. Experimental results are shown in Figure \ref{fig:workload-compara}.

Regarding computational load, \system achieves complete load balancing by partitioning vertex data, as shown in Figure \ref{fig:workload-compara} (a). Regarding total computation, \system is comparable to other full-graph systems but surpasses the DistDGL. Although DistDGL reduces the computation through sampling, the sampling process incurs significant overhead and leads to decreased efficiency as the model layer increases (see details in Section 5.5).
As illustrated in Figure \ref{fig:workload-compara} (b), regarding communication load, \system ensures load balance by assigning each worker an equal number of vertices for embedding \texttt{gather} and \texttt{split}. However, naive tensor parallelism requires embedding \texttt{gather} and \texttt{split} operations at each layer, leading to frequent communication. Benefiting from the decoupled tensor parallelism training approach, \system reduces the overall communication frequency while converting the communication entities into lighter-weight vertex embeddings, significantly reducing communication volume by up to 7.23 $\times$.

\subsection{Performance Gain Analysis of \system}

\begin{figure}
  \centering
  \vspace{-0.1in}
  \includegraphics[width=8cm,page={1}]{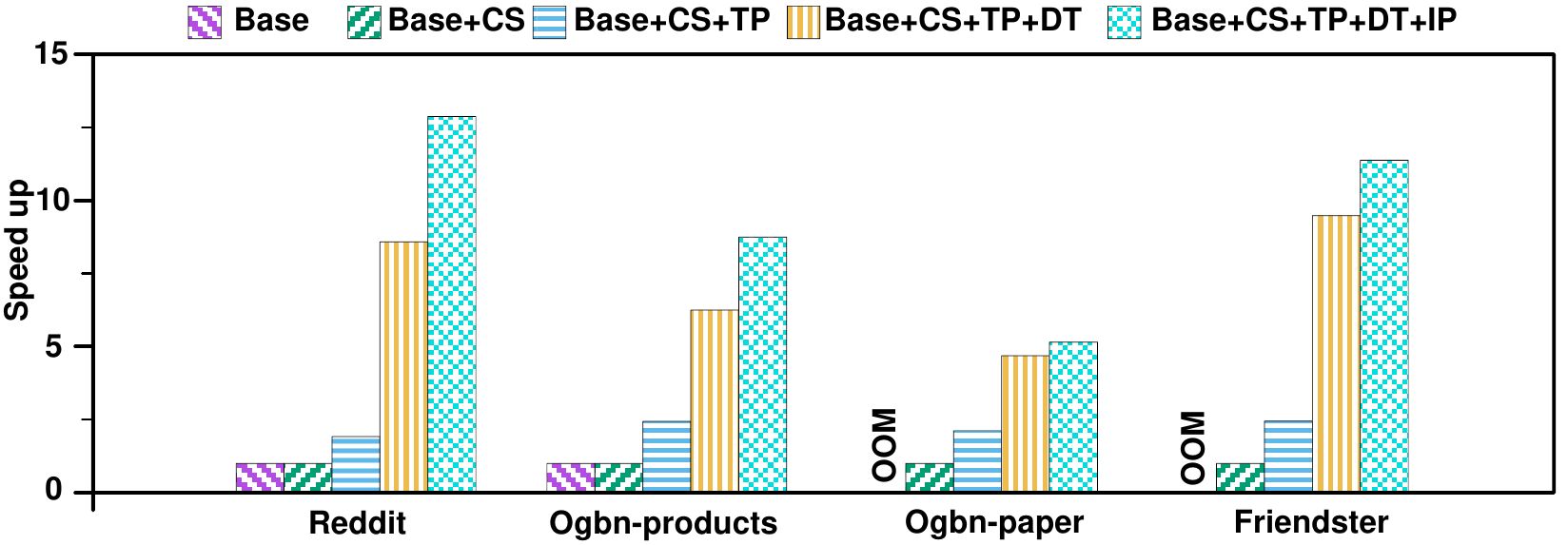}
    \vspace{-0.1in}
% \vspace{-.18in}
  \caption{Performance gain analysis. "CS" indicates the chunk-based task scheduling, "TP" indicates the GNN tensor parallelism training, "DT" indicates the decoupled training method, and "IP" indicates the inter-chunk pipelining.
  }
  \vspace{-0.2in}
  \label{fig:breakdown}
\end{figure}

% \Paragraph{Performance Gain Analysis.}
We analyze the performance gain of GNN tensor parallelism (TP), decoupled training method (DT), chunk-based task scheduling (CS), and inter-chunk pipelining (IP) on the GCN model with four datasets. 
To ensure a fair comparison, we start with a data parallelism baseline based on the \system codebase and gradually integrate the four optimization methods. The data parallelism baseline employs a chunk-based approach for graph partitioning. 
% To mitigate out-of-memory issues in large-scale graphs, all settings are first equipped with chunk-based task scheduling (CS). 
Figure \ref{fig:breakdown} shows the normalized speedups. 
Compared to the baseline, the baseline+CS addresses the memory requirements of large-scale data by further partitioning chunks within each worker. 
% For Reddit and Ogbn-products, distributed training already provides sufficient memory.
Compared to the baseline+CS, TP achieves speedups ranging from 1.92$\times$ to 2.45$\times$ by implementing a more balanced workload. On the Friendster dataset, TP achieves the highest speedup, attributed to its inherent power-law distribution as a social network graph. The chunk-based partitioning strategy exacerbates severe workload imbalances in such graphs. Compared to the baseline+CS+TP, DT achieves speedups ranging from 2.56$\times$ to 4.47$\times$ by significantly reducing communication overhead. DT achieves a 4.47$\times$ speedup on the Reddit dataset, whereas on the Ogbn-paper dataset, the speedup is only 2.21$\times$. This discrepancy is due to the embedding dimension in Reddit being significantly lower than the raw features, facilitating a reduction in communication volume through the early computation of NN operations.
Lastly, IP provides speedups ranging from 1.1$\times$ to 1.5$\times$ by overlapping computation and communication. IP achieves an average speedup of 1.47$\times$ on the Reddit and Product datasets, while on the two larger datasets, the speedup averages only 1.11$\times$. This is because large datasets require more chunks partitioned, leading to more frequent CPU-GPU communication.

%the differing gaps between the raw feature dimensions and the embedding dimensions in these two datasets. T

\subsection{Scalability Analysis}

% 集群扩展性实验
% 实验目的：观察系统的性能随集群规模上升的加速比
% 对比系统：TenparaGNN、NeutronStar、Sancus、DistDGL 
% 数据集：products、reddit
% 实验配置：16节点，T4 GPU
% 模型：GCN
% 实验内容：统计随机器数上升，单个epoch运行时，计算加速比。机器数变化（2-4-8-16）

\begin{figure}
  \centering
  \vspace{-0.1in}
  \includegraphics[width=\linewidth]{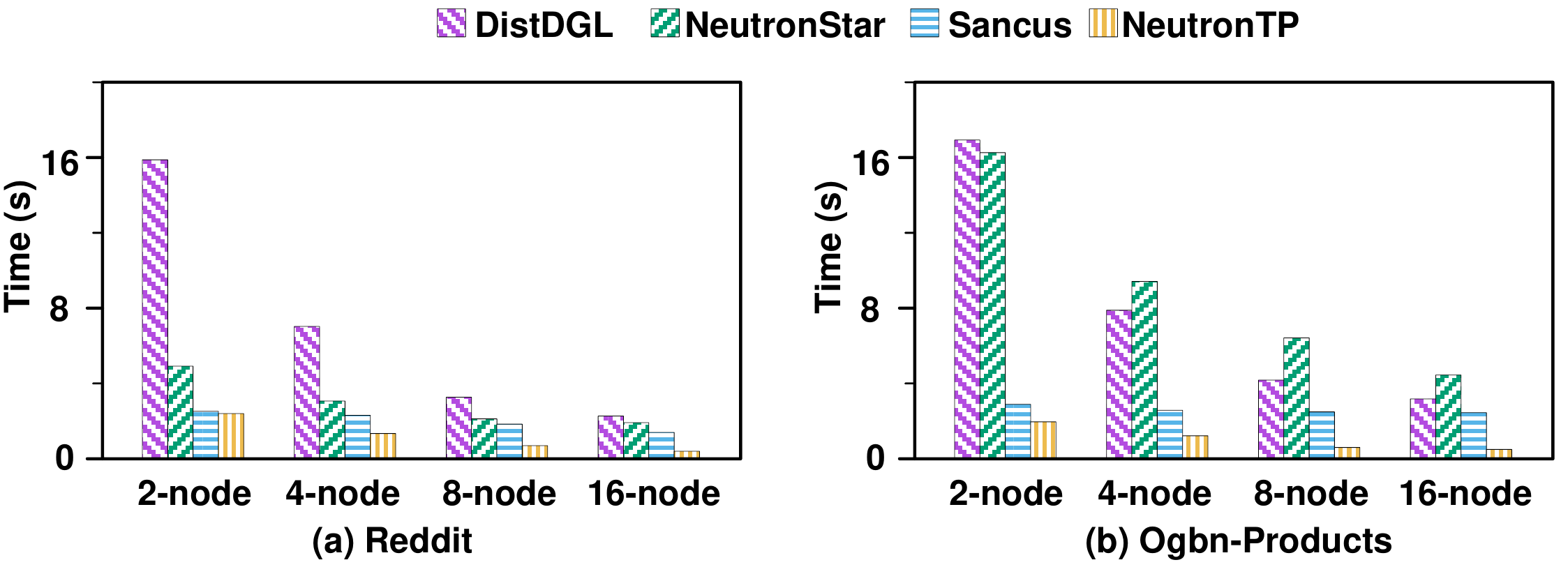}
  \vspace{-0.3in}
  \caption{Per-epoch runtime of different systems with different cluster sizes.}
  \vspace{-0.15in}
  \label{fig:vary-cluster}
\end{figure}

%在本实验中，我们比较了在两个数据集上使用不同集群规模训练GCN时，system与基线的情况。图9展示了逐渐增加集群规模从2-node到16-node，各个系统的性能变化。在不同集群规模下，\system始终表现出优于基线的性能。我们发现\system与其它系统的执行时间随着节点的增加而缩短。然而,Sancus展示了较差的扩展性。这可能与Sancus中的通信实施有关，在Sancus中，每个工作者都需要从远程工作者那里获取整个分区数据，即使只需要一小部分数据。相比之下，\system采用了张量并行消除了顶点依赖同时采用解耦训练方法降低了通信量，并获得了几乎线性的提速。具体来说，当集群内节点的数量从2个增加到16个时，\system分别平均获得了6.33倍，5.97倍和2.69倍的加速比相比于DistDGL，NeutronStar和Sancus
\Paragraph{Performance with varying cluster sizes.}
In this experiment, we compare \system with baselines when training GCN with different cluster sizes over two datasets.
The results are shown in Figure \ref{fig:vary-cluster}. Across different cluster sizes, \system consistently outperforms the baselines. Specifically, as the cluster size increases from 2 to 16, \system achieves an average speedup of 6.33$\times$, 5.97$\times$, and 2.69$\times$ compared to DistDGL, NeutronStar, and Sancus, respectively. We observe that the execution time of \system, DistDGL, and NeutronStar decreases with an increase in the number of nodes. However, Sancus demonstrates poor scalability. That may be due to its communication implementation, where each worker needs to fetch the entire partition data from remote workers, even if only a small portion of the data is required. In contrast, \system adopts tensor parallelism to eliminate vertex dependencies and employs a decoupled training approach to reduce communication overhead, achieving nearly linear speedup. Specifically, as the number of nodes in the cluster increases from 2 to 16, \system achieves an average speedup of 6.33$\times$ and 4.97$\times$ on Reddit and Ogbn-products.

% , and 2.69$\times$ compared to DistDGL, NeutronStar, and Sancus, respectively.

% 模型层数扩展性实验
% 实验目的：观察系统的性能随模型层数加深的变化情况
% 对比系统：TenparaGNN、NeutronStar、Sancus、DistDGL 
% 数据集：products、reddit
% 实验配置：16节点，T4 GPU
% 模型：GCN
% 实验内容：统计随模型层数上升，单个epoch运行时。模型层变化（2-3-4）

\begin{figure}
% \vspace{-0.4in}
  \centering
  \includegraphics[width=\linewidth]{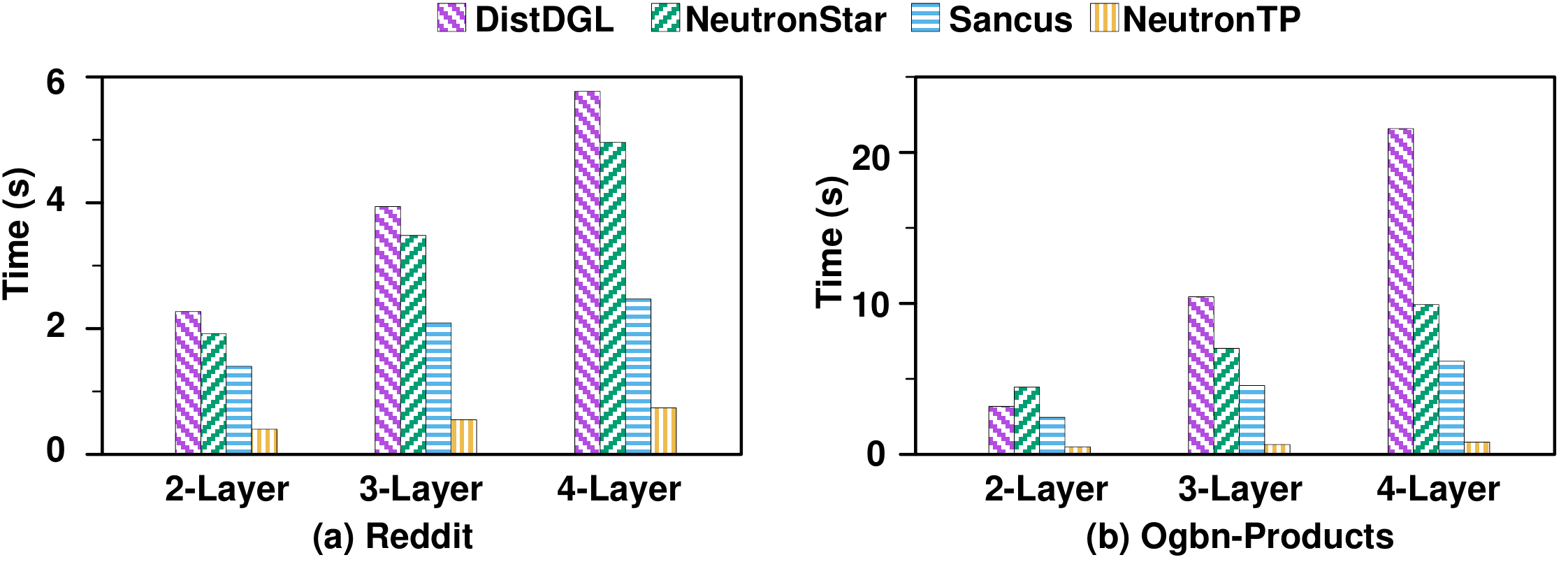}
  \vspace{-0.3in}
  \caption{Per-epoch runtime of different systems with different model layers.}
  \label{fig:vary-layer}
  \vspace{-0.15in}
\end{figure}

%在本实验中，我们比较了在 16 节点集群中通过两个数据集训练具有不同模型深度的 GCN 时，system 与基线的情况。我们为 GNN 模型创建了三种变体，分别为 2 层、3 层和 4 层。对于 2 层、3 层和 4 层模型，DGL 的采样策略分别为（25，10）、（25，15，10）和（25，20，15，10）。实验结果如图9所示。我们观察到\system相比于其他系统的性能优势随着模型深度增加而逐渐增加。对于2-layer模型，\system平均获得了5.99倍的加速比。对于3-layer模型与4-layer模型，加速比分别是8.65倍与11.13倍。正如我们在section 2中描述的，基于数据并行的分布式GNN训练系统的跨worker顶点依赖会随着模型深度增加而显著上升。DistDGL的效率下降最严重因为它的采样策略会导致每一层的模型规模呈指数上升，而full-graph系统的模型规模只会线性增加。具体地，\system超过DistDGL最高26.64倍在ogbn-product数据上的四层模型中。这是因为Ogbn-product的平均度数远小于Reddit，图拓扑更加稀疏，导致采样子图规模上升更快。\system相比于其他baseline的显著提升说明了张量并行通过消除顶点依赖在处理更深层模型时保持了高效。
\Paragraph{Performance with varying model layers.}
In this experiment, we compare \system with baselines when training GCN with different model layers over two datasets in a 16-node cluster. For the 2, 3, and 4-layer models, the DGL sampling strategies were set to (25,10), (25,15,10), and (25,20,15,10) respectively. The results are shown in Figure \ref{fig:vary-layer}. We observe that the performance advantage of \system over other baselines gradually increased with the model depths. For the 2-layer model, \system achieves an average speedup of 5.99$\times$. For the 3-layer and 4-layer models, the speedups were 8.65$\times$ and 11.13$\times$ respectively. 
\notecolor{This is because \system's tensor parallelism can effectively eliminate the cross-worker vertex dependencies among layers in GNN data parallel training, thereby removing substantial communication overhead.}
% As described in Section \ref{sec2}, the cross-worker vertex dependencies in data parallelism GNN training significantly increase with the model layers.
DistDGL experiences the most severe efficiency degradation because of the neighbor explosion problem \cite{ROC_mlsys_2020}, where the computation and memory requirements for mini-batch GNN training increase exponentially with the number of layers.
% In contrast, the model size of full-graph systems only increases linearly.
\system outperforms DistDGL by up to 26.64$\times$ in the 4-layer model on the Ogbn-products dataset. This is because the average degree of Ogbn-products is much smaller than that of Reddit, making the graph topology sparser and causing a faster increase in the sampled subgraph size. 
% The significant improvement of \system over other data parallelism baselines demonstrates the effectiveness of tensor parallelism in handling deeper models by eliminating vertex dependencies.

\begin{figure}
\vspace{-0.1in}
  \centering
  \includegraphics[width=\linewidth]{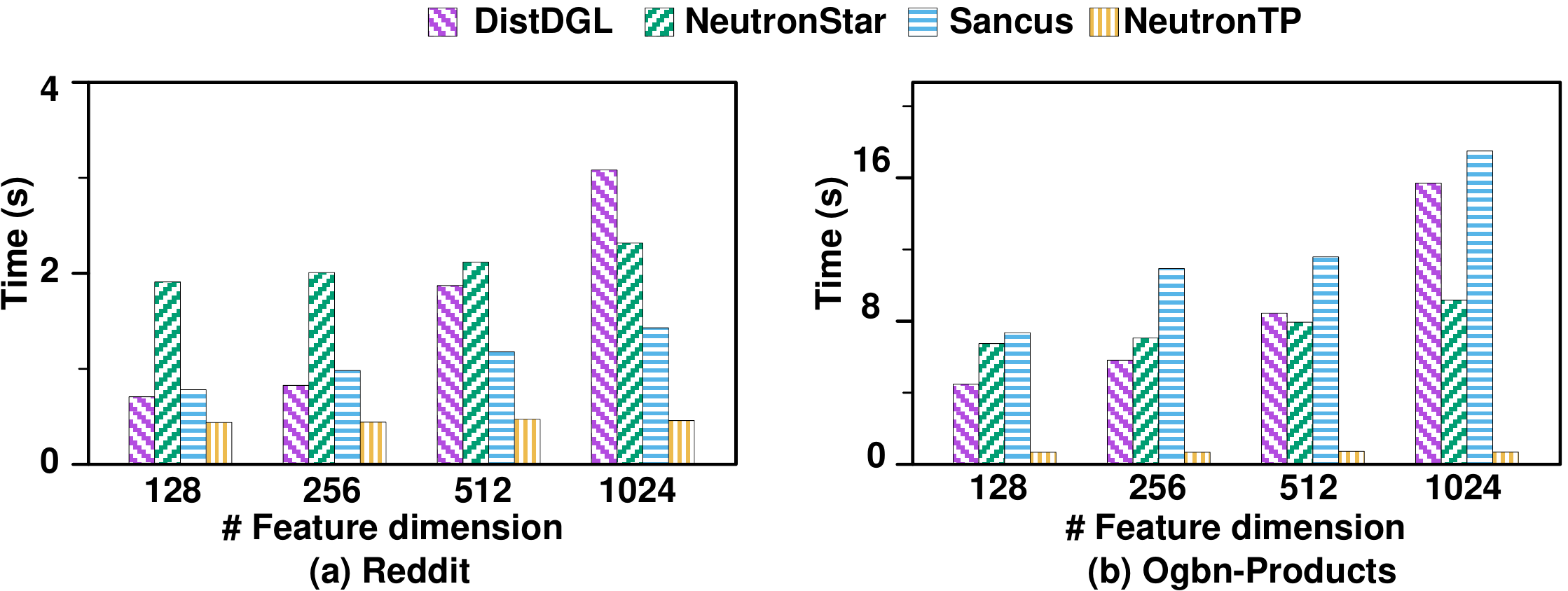}
  \vspace{-0.3in}
  \caption{\notecolor{Per-epoch runtime of different systems with different feature dimensions.}}
  \label{fig:vary-feature}
  \vspace{-0.1in}
\end{figure}

\Paragraph{\notecolor{Performance with varying feature dimensions.}} \notecolor{In this experiment, we compare \system with baselines when training GCN with different input feature dimensions over two datasets in a 16-node cluster. The results are shown in Figure \ref{fig:vary-feature}. We observe that the performance advantage of NeutronTP over other baselines gradually increased with the feature dimensions. For the 128-dimensional dataset, \system achieves an average speedup of 5,87$\times$. For the 256-dimensional, 512-dimensional, and 1024-dimensional datasets, the speedups are 7.28$\times$, 8.14$\times$, and 12.74$\times$ respectively. When feature dimensions increase, GNN data parallelism suffers from significant communication overhead, particularly during the communication of raw features in the first layer. NeutronTP employs decoupled GNN tensor parallelism, which only gathers and splits vertex embeddings. Compared to GNN data parallelism, it reduces communication frequency and transforms features into lower-dimensional embeddings before communication.}

% \subsection{Complex Model Support}

% 资源利用率对比实验
% 实验目的：观察系统对算力与带宽资源的利用情况
% 对比系统：TenparaGNN、NeutronStar、Sancus、DistDGL 
% 数据集：products、reddit
% 实验配置：16节点，T4 GPU
% 模型：GCN
% 实验内容：统计CPU，GPU利用率，通信带宽利用率（单位时间通信量除带宽）。

%我们在Reddit上对NeutronTP和基线的GCN训练过程中评估了GPU的利用率。图13显示了20秒时间窗口内的结果。GPU利用率每100毫秒记录一次，并以1秒为间隔取平均值。NeutronTP获得更高的平均GPU利用率相比于DistDGL，Sancus和NeutronStar，并在大部分时间内拥有更高的GPU利用率峰值。DistDGL的GPU利用率相当低。这是因为DistDGL有一个涉及大量随机访问的采样步骤，这可能是限制GPU利用率的瓶颈。另外，基于GNN数据并行的baselines都会因为不平衡的工作负载而导致部分worker处于闲置状态，降低总体GPU利用率。NeutronTP最小化GPU空闲时间，归功于其平衡的工作负载和重叠通信与计算的inter-sugraph的流水线设计。通过利用GNN张量并行来保证均衡的工作负载，最大程度避免了GPU闲置的问题，并通过子图之间的流水线设计，最大限度地增加了通信和计算的重叠。

\begin{figure}
  \centering
   \vspace{-0.1in}
  \includegraphics[width=0.9\linewidth,page={1}]{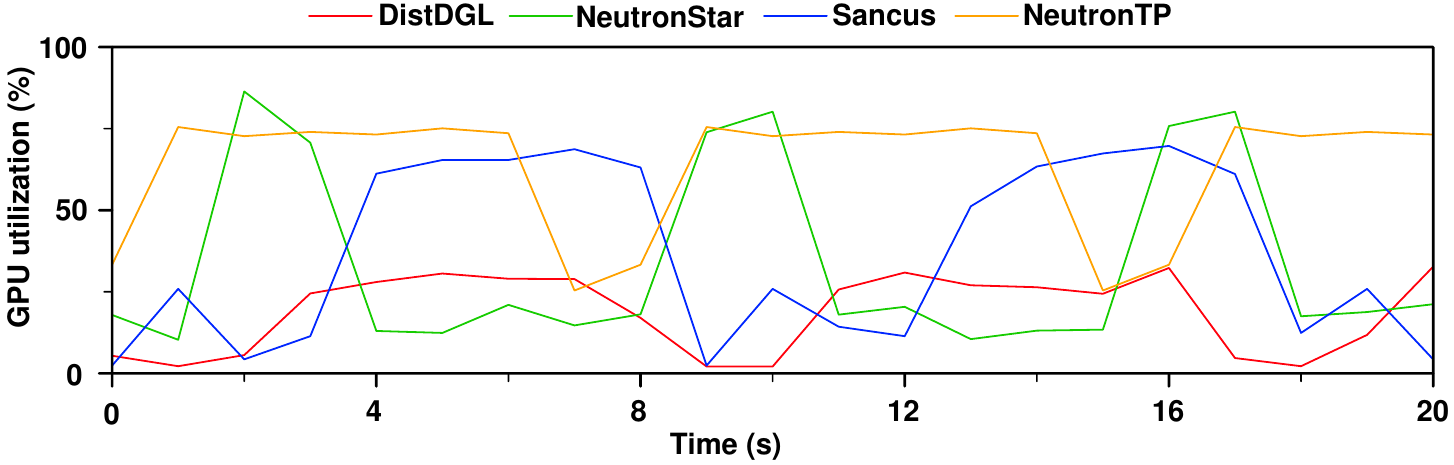}
    \vspace{-0.15in}
  \caption{GPU utilization comparison. The average GPU utilizations are 62.85\%, 19.91\%, 33.97\%, and 37.67\% for \system, DistDGL, NeutronStar, and Sancus, respectively.}
  \label{fig:uti}
   \vspace{-0.1in}
\end{figure}

\subsection{GPU Utilization}
We evaluate the GPU utilization during the training of GCN on Reddit for \system and baselines in a 16-node cluster. Figure \ref{fig:uti} shows the results in a 20-second time window. The GPU utilization is recorded every 100 milliseconds and averaged in a 1-second interval. \system exhibits higher GPU utilization (62.85\% on average) compared to DistDGL (19.91\% on average), Sancus (37.67\% on average), and NeutronStar (33.97\% on average), with consistently higher peak GPU utilization for most of the time. 
DistDGL shows relatively low GPU utilization due to the sampling step involving a large amount of random access, which can be the bottleneck to limit GPU utilization. Additionally, baselines based on GNN data parallelism suffer from GPU idle time due to unbalanced workloads, resulting in decreased overall GPU utilization. \system experiences minimal GPU idle time, attributed to balanced workloads and a pipeline design between chunks, maximizing the overlap of communication and computation.

\label{acc}
\subsection{Accuracy Comparisons}
\begin{figure}
  \centering
   \vspace{-0.1in}
  \includegraphics[width=0.9\linewidth,page={1}]{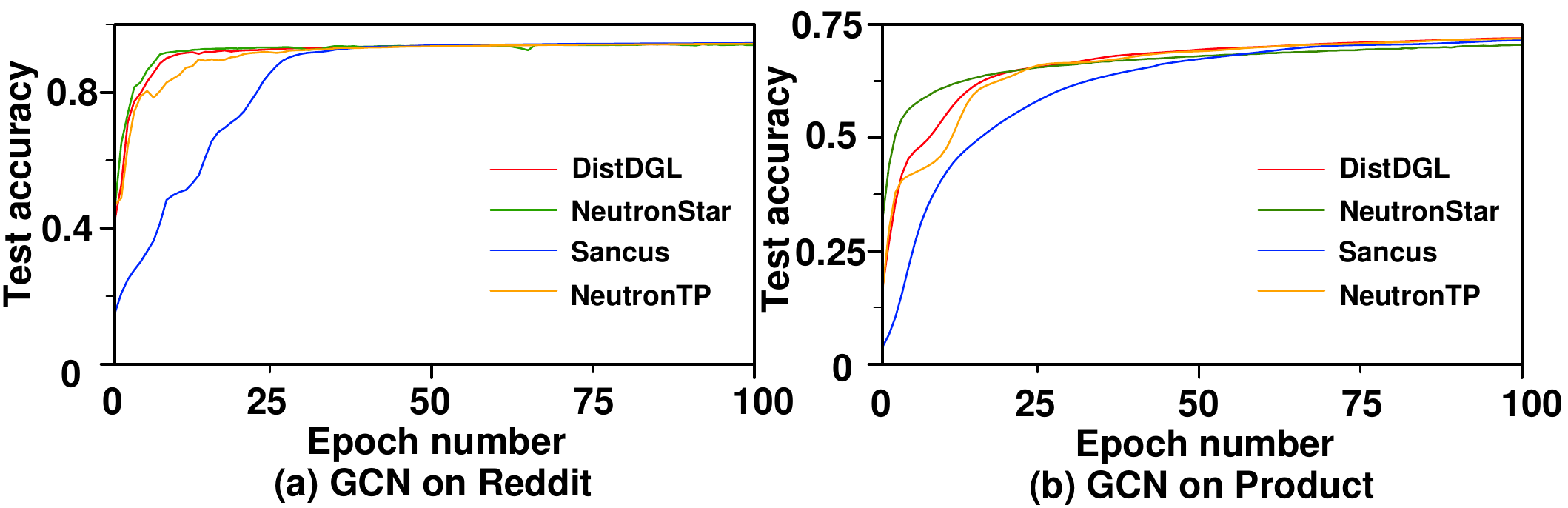}
  \vspace{-0.1in}
  \caption{Epoch-to-accuracy.}
   \vspace{-0.2in}
  \label{fig:acc}
\end{figure}

\notecolor{The changes in the execution process introduced by the decoupled training method may impact model performance. We plot the epoch-to-accuracy curve on different systems for a GCN model over two datasets. The results are shown in Figure \ref{fig:acc}. After 100 epochs, the test accuracy reaches a stable state, \system and other baselines almost achieve the same test accuracy. However, NeutronTP converges slightly slower compared to the traditional GNN training method.} 
% We plot the epoch-to-accuracy curve on different systems for a GCN model over two datasets. The results are shown in Figure \ref{fig:acc}. After 100 epochs, the test accuracy reaches a stable state, \system and other baselines almost achieve the same test accuracy, demonstrating the efficacy of decoupled GNN training in ensuring model quality. 
Sancus exhibits the slowest increase in accuracy over epochs due to its use of historical embeddings. Additionally, it is worth noting that while many works \cite{ROC_mlsys_2020,distgnn_sc_2021} have demonstrated that sampling strategies can lead to lower accuracy, we find that DistDGL performs close to full-graph training accuracy on commonly used datasets. We attribute this to the well-tuned parameters of DistDGL. \notecolor{Both mini-batch and full-graph training methods have their advantages. In summary, NeutronTP exhibits advantages when training deep GNNs or when the input graph includes a large proportion of training vertices. However, when the training set and the number of model layers are small, DistDGL still holds certain advantages.}

% Both mini-batch and full-graph training methods have their advantages. However, evaluating their impact on accuracy requires a comprehensive analysis and consideration of various factors, including batch size, sampling fanout, and characteristics of the input graph, which is out of the scope of this work.

\subsection{\notecolor{Extension to heterogeneous graphs}}
\notecolor{NeutronTP can be naturally extended to heterogeneous graphs. We compare NeutronTP with DistDGLv2 \cite{dglv2_kdd_2022} using two heterogeneous graphs \cite{ogblsc_nips_2021,ogb_2020_neurIPS} and the R-GCN algorithm \cite{RGCN_ESWC_2018} in a 16-node cluster. DistDGLv2 extends DistDGL to support heterogeneous GNN training. R-GCN are designed to handle heterogeneous graphs, which consist of multiple types of edges. On the Ogbn-mag dataset, NeutronTP achieves a 6.15$\times$ speedup. On the Mag-lsc dataset, DistDGLv2 exhibits better performance as it trains on only 0.4\% of the total vertices, reducing the computational load compared to NeutronTP. This is similar to the Ogbn-paper dataset, which includes only 1.1\% training vertices, leading to significant differences in overall computation between full-graph and mini-batch training. }
% However, from a throughput perspective, NeutronTP maintains a significant advantage on the Mag-lsc dataset, achieving an average of 351K vertices/s compared to 19.3K vertices/s for DistDGLv2.}

\subsection{\notecolor{Training cost breakdown}}
\notecolor{We evaluate the training costs of different stages for both node classification and link prediction tasks using the GCN algorithm with the Reddit dataset. The results are presented in Table \ref{tab:downstream task}. We can observe that GNN computation is the primary cost, accounting for an average of 94\% of the time in node classification and 79\% in link prediction.  By optimizing the GNN computation process, NeutronTP reduces end-to-end training time by 65\% to 80\% compared to NeutronStar. Therefore, the practical downstream tasks that benefit most from NeutronTP are those where GNN computation is a significant part of the overall training cost.}

% \subsection\notecolor{\textbf{Section 5.9 Training cost breakdown} We evaluate the training costs of different stages for both node classification and link prediction tasks using GCN algorithm with the Reddit dataset. The results are presented in Table 4. We can observe that GNN computation is the primary cost, accounting for an average of 95\% of the time in node classification and 83\% in link prediction.  By optimizing GNN computation process, NeutronTP reduces end-to-end training time by 65\% to 80\% compared to NeutronStar, enhancing performance in both tasks. Therefore, the practical downstream tasks that benefit most from NeutronTP are those where GNN computation is a significant part of the overall training cost.} 
\section{Related work}
\label{sec6}

\Paragraph{Full-graph GNN training.} A set of GNN systems \cite{neugraph_atc_2019,neutron_sigmod_2022,DGCL_eurosys_2021,ROC_mlsys_2020,G3_SIGMOD_2023,SANCUS_VLDB_2022,hongtu_SIGMOD_2024,PipeGCN_iclr_2022,CAGNET_2020_SC,PYG_2019_arxiv} adopt full-graph GNN training to guarantee high accuracy. NeuGraph \cite{neugraph_atc_2019} defines a new, flexible SAGA-NN model to express GNNs. 
% It fuses graph operations into TensorFlow and supports training on multiple GPUs.
CAGNET \cite{CAGNET_2020_SC} proposes 1.5D, 2D, and 3D graph partitioning to optimize the data distribution among GPUs.
ROC \cite{ROC_mlsys_2020} optimizes graph partitioning with online learning while managing memory with dynamic programming. 
DGCL \cite{DGCL_eurosys_2021} designs a communication planning algorithm that avoids conflict on various links. 
PipeGCN \cite{PipeGCN_iclr_2022} reduces communication for boundary vertices by utilizing historical embeddings and effectively overlaps computation across different partitions.
G3 \cite{G3_SIGMOD_2023} propose GNN hybrid parallelism to scale out GNN training with carefully scheduled peer-to-peer intermediate data sharing. Hongtu \cite{hongtu_SIGMOD_2024} designs a recomputation-cache-hybrid intermediate data management to significantly reduce the GPU memory requirement.

%pipegcn通过使用历史嵌入减少边界点的通信并充分重叠不同分区之间的计算过程。

%GNN模型中复杂的图拓扑结构和各种工作量特征（模型深度，batchsize，训练点分布等等）增加了在工人之间平衡划分工作量的难度。许多GNN训练的综述文章都将负载不均衡归纳为主要的挑战，无论是minibatch训练还是fullgraph训练。另外，一些实验文章也通过大量实验分析证明了负载不均在分布式GNN训练中普遍存在,并可能与其他划分目标冲突如最小化通信。许多GNN系统探索适应GNN工作负载的图划分策略来避免负载不均衡问题。SALIENT++采用扩展的METIS划分，在最小的边割的前提下，加入额外的约束条件以尽可能实现负载平衡。PaGraph与ByteGNN采用流式划分方法来细粒度的划分图数据，在划分过程中为每个顶点逐个选择最佳分区。G3采用迭代分区方法，在不同分区之间不断交换顶点以尽可能实现负载平衡。考虑到真实世界图数据的复杂性，这些图划分算法通常只能逼近负载均衡状态，且划分效果可能随着图数据的幂律性升高而降低。另外，这些方法的内存开销与时间开销同样不可忽视，尤其是METIS与流式划分，他们的开销甚至可能超过训练本身。因此，我们相信GNN张量并行划分特征而不是划分图数据是一个值得应用的分布式训练方法。
\Paragraph{Load balancing study in distributed GNN training.}
% The complex graph topology and various workload characteristics (such as model depth, batch size, training vertex distribution, etc.) in GNN models increase the difficulty of balancing workload partitioning.
Many survey papers \cite{distgnn_survey_1,distgnn_survey_2,pdgnn_2024_tpami} highlight workload imbalance as a primary challenge in distributed GNN training. 
Experimental studies \cite{exp_vldb_2024,exp-par_vldb_2024} have empirically demonstrated the prevalence of load imbalance.
%and may conflict with other partitioning objectives such as minimizing edge-cuts. 
Many GNN systems attempt to address this issue by exploring graph partitioning strategies \cite{SALIENTPlus_MLSYS23,CAGNET_2020_SC,PaGraph_SoCC20,ByteGNN_VLDB22,G3_SIGMOD_2023,distdgl_sc_2020}. 
% Both DistDGL \cite{distdgl_sc_2020} and SALIENT++ \cite{SALIENTPlus_MLSYS23} utilize the METIS partitioning method. METIS ensures minimal edge cuts initially and can incorporate additional constraints to achieve load balance as much as possible, but this comes at the cost of increased time and memory overhead.
SALIENT++ \cite{SALIENTPlus_MLSYS23} extends the METIS partitioning approach with additional constraints to balance workloads while minimizing edge-cuts. 
PaGraph \cite{PaGraph_SoCC20} and ByteGNN \cite{ByteGNN_VLDB22} employ streaming partitioning methods, selecting the optimal partition for each vertex individually. G3 \cite{G3_SIGMOD_2023} adopts an iterative partitioning approach, continuously exchanging vertices between different partitions. 
Given the complexity of real-world graph data, these methods often approximate load balance and may experience diminishing effectiveness with increasing graph power-law characteristics \cite{exp_vldb_2024,exp-par_vldb_2024}. Furthermore, the memory and computation overhead of these methods is considerable, especially for METIS and streaming partitioning, which may even surpass the training itself \cite{exp_vldb_2024,info24,P3_OSDI_2021}. 
Therefore, we believe that partitioning features instead of graph data for GNN tensor parallelism is a promising distributed training approach.

\Paragraph{Vertical feature partitioning.}
Some recent studies \cite{P3_OSDI_2021,info24} explore vertical feature partitioning in distributed GNN training.
P3 \cite{P3_OSDI_2021} utilizes feature slices to complete the first graph aggregation operation, reducing feature fetching overhead. Du et al. \cite{info24} propose skipping feature fetching in some iterations, leveraging only partial feature dimensions for local training to achieve a trade-off between convergence error and feature communication time.

%我们提出了\system, 一个负载均衡且高效的分布式全图GNN训练系统。\system使用GNN张量并行展开分布式训练，在不同worker间划分规则的特征而不是复杂的图结构。相比于GNN数据并行，\system消除了跨worker的顶点依赖并且获得了充分的负载均衡。\system克服了GNN张量并行的独特挑战，它使用解耦的训练方法来显著降低通信，并且使用内存高效的子图调度策略来降低内存开销同时重叠计算与通信。大量实验表明，与GNN数据并行相比，我们的方法大大加快了分布式 GNN 训练的速度，同时实现了相当水平的模型准确性和收敛速度。

\begin{table}
        \centering
        \vspace{-0.1in}
    \caption{\notecolor{Comparison with DistDGLv2 on two heterogeneous graphs.}}
    \vspace{-0.2in}
    \label{tab:heter data}
    \footnotesize
	\centering
	{\renewcommand{\arraystretch}{1.2}
	\begin{tabular}{l c c}
		\hline
		\multirow{2}*{\textbf{System}}  & \multicolumn{2}{c}{Runtime of R-GCN (s)}\\
        \cline{2-3}
        &\multirow{1}*{Ogbn-mag}  & 
        \multirow{1}*{Mag-lsc}  \\
		\hline
        {DistDGLv2}  & {36.3} & \textbf{56.9} \\ 
    	{NeutronTP} & \textbf{5.9} & {695.2} \\  
		\hline
	
		\hline
	\end{tabular}
	}
 \vspace{-0.3cm}
\end{table}

\begin{table}
        \centering
    \caption{\notecolor{The runtime breakdown (in seconds) with different tasks. (NC: node classification, LP: link prediction)}}
     \vspace{-0.1in}
    \label{tab:downstream task}
    \footnotesize
    % \small
	\centering
	\renewcommand{\arraystretch}{1.2}
 \resizebox{0.45\textwidth}{!}{
	\begin{tabular}{l c c c c c c}
		\hline
        \multirow{2}*{\textbf{Task}}  &
		\multirow{2}*{\textbf{System}}  & 
        \multirow{1}*{\textbf{Negative }}  & 
        \multirow{1}*{\textbf{GNN }}  &
        \multirow{1}*{\textbf{Classification}}  &
        \multirow{1}*{\textbf{Loss}}  & \\
        & &\multirow{1}*{\textbf{Sampling}}  & 
        \multirow{1}*{\textbf{Computation}}  &
        \multirow{1}*{\textbf{Computation}}  &
        \multirow{1}*{\textbf{Calculation}} \\
		\hline
        \multirow{2}* {NC} & {NeutronStar}  & {-/-} & \textbf{1.88/97\%} & {0.03/2\%} & {0.01/1\%} \\ 
    	& {NeutronTP} & {-/-} & \textbf{0.36/90\%} & {0.03/7\%} & {0.01/3\%}\\  
     \hline
        \multirow{2}*{LP}& {NeutronStar}   & {0.07/3\%} & \textbf{2.12/90\%} & {0.11/5\%} & {0.04/2\%} \\ 
    	 & {NeutronTP} & {0.07/9\%} & \textbf{0.53/67\%} & {0.15/19\%} & {0.04/5\%}\\ 
		\hline
	
		\hline
	\end{tabular}
	}
 \vspace{-0.1in}
\end{table}

\section{Conclusion}
\label{sec7}

We present \system, a load-balanced and efficient distributed full-graph GNN training system. \system leverages GNN tensor parallelism for distributed training, which partitions feature rather than graph structures. Compared to GNN data parallelism, \system eliminates cross-worker vertex dependencies and achieves a balanced workload. To address the unique challenges of GNN tensor parallelism, \system employs a generalized decoupled training approach to significantly reduce communication overhead and a memory-efficient task scheduling strategy to reduce memory consumption while overlapping computation and communication. Extensive experiments demonstrate that our approach accelerates distributed GNN training significantly compared to GNN data parallelism while achieving comparable model accuracy.
%\clearpage

% \vspace{-0.10in}
\begin{acks}
% We thank the anonymous reviewers for their constructive comments and suggestions.
This work is supported by the National Natural Science Foundation of China (U2241212, 62072082, 62202088, 62072083, and 62372097), the 111 Project (B16009), and the Distinguished Youth Foundation of Liaoning Province (2024021148-JH3/501). Yanfeng Zhang and Qiange Wang are the corresponding authors.
\end{acks}

\bibliographystyle{ACM-Reference-Format}
\bibliography{sample}

\end{document}